\documentclass[12pt]{article}
\date{August 2, 2016}
\usepackage[usenames,dvipsnames]{color}
\usepackage{amsmath,amsthm,amsfonts,amssymb,multicol,amscd,amsbsy}
\usepackage{graphicx}
\usepackage{booktabs}
\usepackage{array}
\usepackage{subfigure}
\usepackage{enumerate}
\usepackage{url}
\usepackage[nodayofweek]{datetime}
\usepackage[english]{babel}
\usepackage{mathtools}
\usepackage[toc,page]{appendix}
\usepackage{verbatim} 
\usepackage{amsthm} 
\usepackage{enumitem}
\usepackage{enumerate}
\usepackage{bbm} 
\usepackage[normalem]{ulem} 
\usepackage{bm}
\usepackage{hyperref}

\usepackage{fullpage}
\usepackage{authblk}

\newcommand{\be}{\begin}
\newcommand{\e}{\end}
\newcommand{\beq}{\begin{equation}}
\newcommand{\eeq}{\end{equation}}
\newcommand{\beqs}{\begin{equation*}}
\newcommand{\eeqs}{\end{equation*}}
\newcommand{\bal}{\begin{align}}
\newcommand{\eal}{\end{align}}
\newcommand{\bals}{\begin{align*}}
\newcommand{\eals}{\end{align*}}

\newcommand{\ol}{\overline}

\newcommand{\x}{\mathbf{x}}

\renewcommand{\l}{\left}
\renewcommand{\r}{\right}

\renewcommand{\d}{\mathrm{d}} 
\newcommand{\De}{\Delta}
\renewcommand{\Re}{\mathrm{Re}}

\newcommand{\set}[1]{\mathbb{#1}}

\newcommand{\curly}[1]{\mathcal{#1}}
\newcommand{\goth}[1]{\mathfrak{#1}}
\newcommand{\setof}[2]{\left\{ #1\; : \;#2 \right\}}

\newcommand{\R}{\set{R}}

\newcommand{\D}{\curly{D}}

\newcommand{\Om}{\Omega}
\newcommand{\eps}{\epsilon}

\newcommand{\lam}{\lambda}

\newcommand{\gam}{\gamma}
\newcommand{\Gam}{\Gamma}

\newcommand{\al}{\alpha}
\newcommand{\de}{\delta}

\newcommand{\vp}{\varphi}

\newcommand{\dist}{\mathrm{dist}}

\newcommand{\ttmatrix}[4]{\left(\be{array}{cc} #1&#2\\	#3&#4 \e{array}	\right)}

\newcommand{\scpp}[2]{\l\langle#1,#2\r\rangle}
\newcommand{\qmscp}[3]{\langle #1 \l\vert #2 \r\vert #3 \rangle}

\renewcommand{\it}{\infty}

\newcommand{\supp}{\,\mathrm{supp}\,}

\newcommand{\del}{\partial}

\newcommand{\Tr}[1]{\textnormal{Tr}\left[#1\right]}	
\newcommand{\tr}[1]{\textnormal{tr}\left[#1\right]}	
\newcommand{\ft}[1]{\widehat{{#1}}}			

\theoremstyle{definition}

\numberwithin{equation}{section}

\theoremstyle{remark}

\def\dotuline{\bgroup
  \ifdim\ULdepth=\maxdimen  
   \settodepth\ULdepth{(j}\advance\ULdepth.4pt\fi
  \markoverwith{\begingroup
  \advance\ULdepth0.08ex
  \lower\ULdepth\hbox{\kern.15em .\kern.1em}%
  \endgroup}\ULon}

\def\dashuline{\bgroup
  \ifdim\ULdepth=\maxdimen  
   \settodepth\ULdepth{(j}\advance\ULdepth.4pt\fi
  \markoverwith{\kern.15em
  \vtop{\kern\ULdepth \hrule width .3em}%
  \kern.15em}\ULon}
\allowdisplaybreaks

\begin{document}
\title{Condensation of fermion pairs in a domain}

\author{Rupert L. Frank\thanks{rlfrank@caltech.edu} }
\author{Marius Lemm\thanks{mlemm@caltech.edu} }
\author{Barry Simon\thanks{bsimon@caltech.edu}} 
\affil{Mathematics Department, Caltech}


\renewcommand\Authands{ and }

\maketitle

\abstract{
We consider a gas of fermions at zero temperature and low density, interacting via a microscopic two body potential which admits a bound state. The particles are confined to a domain with Dirichlet (i.e.\ zero) boundary conditions. Starting from the microscopic BCS theory, we derive an effective macroscopic Gross-Pitaevskii (GP) theory describing the condensate of fermion pairs. The GP theory also has Dirichlet boundary conditions.

Along the way, we prove that the GP energy, defined with Dirichlet boundary conditions on a bounded Lipschitz domain, is continuous under interior and exterior approximations of that domain.
}

\setcounter{tocdepth}{1} 
\tableofcontents

\section{Introduction}

We consider a gas of fermions at zero temperature in $d=1,2,3$ dimensions and at chemical potential $\mu<0$. The particles are confined to an open and bounded domain $\Omega\subseteq\R^d$ with Dirichlet (i.e.\ zero) boundary conditions. They interact via a microscopic local two body potential $V$ which admits a two body bound state. Additionally, the particles are subjected to a weak external field $W$, which varies on a macroscopic length scale.

At low particle density, this leads to tightly bound fermion pairs. The pairs will approximately look like bosons to one another and, since we are at zero temperature, they will form a Bose-Einstein condensate (BEC). It was realized in the 1980s \cite{Leggett79, NS85} that BCS theory, initially used to describe Cooper pair formation in superconductors on much larger (but still microscopic) length scales \cite{BCS57}, also applies in this situation. Moreover, the macroscopic variations of the condensate density are given in terms of the nonlinear Gross-Pitaevskii (GP) theory \cite{DZ92, PS03, SRE93}. An effective GP theory was recently derived mathematically starting from the microscopic BCS theory, see \cite{B, HS12} for the stationary case and \cite{HS13} for the time-dependent case. This is in the spirit of Gorkov's paper \cite{Gorkov} on how Ginzburg-Landau theory arises from BCS theory for superconductors \emph{at positive temperature}. The latter problem has been intensely studied mathematically in recent years \cite{FHSS12, FHSS12c, FHSS15, FL15, HSreview}.

The papers mentioned above all work under the assumption that the system has no boundary (either by working on the torus or on the whole space). In the present paper, we start from low-density BCS theory with Dirichlet boundary conditions and we show that the effective macroscopic GP theory \emph{also has Dirichlet boundary conditions}. 

Our result is also new in the linear setting. The formal statement and its comparatively short proof can be found in Appendix \ref{app:linear} and we hope that this part may serve to illustrate the ideas to a wider audience. In a nutshell, in the linear case we consider the two body Schr\"odinger operator
$$
H_h:=\frac{h^2}{2}(-\De_{\Om,x}+W(x)-\De_{\Om,y}+W(y))+V\l(\frac{x-y}{h}\r),
$$
acting on $L^2(\Om\times\Om)$, where $-\De_\Om$ is the Dirichlet Laplacian. $H_h$ describes the energy of a fermion pair confined to $\Om$. While the center of mass variable $\frac{x+y}{2}$ and the relative variable $x-y$ \emph{do not decouple as usual due to the boundary conditions}, we show that the ground state energy of $H_h$ can be computed in a decoupled manner when $h\to 0$. Namely, one can separately minimize (a) in the relative variable without boundary conditions and (b) in the center of mass variable with Dirichlet boundary conditions and combine the results to obtain the leading and subleading terms in the asymptotics ground state energy of $H_h$ as $h\downarrow 0$. For the details, we refer to Theorem \ref{thm:twobody}.

At positive temperature, de Gennes \cite{deGennesBC} predicted that BCS theory with Dirichlet boundary conditions should instead lead to a Ginzburg-Landau theory with Neumann boundary conditions. We believe that the discrepancy with our result here is due to the fact that we study the system in the low density limit.

\subsection{BCS theory with a boundary}
Let $\Om\subset \R^d$, $d=1,2,3$ be open, further assumptions on $\Om$ are described below. In the BCS model, one restricts to \emph{BCS states} (also called ``quasi-free'' states), which are fully described by an operator
\beq
\label{eq:Gamdefn}
\Gam=\ttmatrix{\gam}{\al}{\ol{\al}}{1-\ol{\gam}},\qquad 0\leq \Gam\leq 1
\eeq
acting on $L^2(\Om)\oplus L^2(\Om)$. Physically, $\gam$ is the one body density matrix and $\al$ is the fermion pairing function, see also Remark \ref{rmk:BCS} (ii). The condition $0\leq \Gam\leq 1$ implies that $0\leq \gam\leq 1$, $\ol{\al}=\al^*$ and $0\leq \al\ol{\al}\leq \gam-\gam^2$. 

 We let $h>0$ denote the (small) ratio between the microscopic and macroscopic length scales. The energy of unpaired electrons at chemical potential $\mu<0$ is described by the one body Hamiltonian
$$
\goth{h}=-h^2\Delta_\Om+h^2W-\mu,\qquad W:\Om\to \R.
$$
Here, $-\Delta_\Om$ is the Dirichlet Laplacian on $\Om$. By definition, it is the self-adjoint operator which by the KLMN theorem corresponds to the quadratic form
$$
\int_\Om |\nabla f(x)|^2 \d x,\qquad f\in H_0^1(\Om).
$$

In macroscopic units, the \emph{BCS energy} of a BCS state $\Gam$ is given by
\beq
\label{eq:EBCSdefn}
\curly{E}_\mu^{BCS}(\Gam)=\Tr{\goth{h} \gam}+\iint\limits_{\Om^2}V\l(\frac{x-y}{h}\r) |\al(x,y)|^2\d x \d y.
\eeq

\be{rmk}
\label{rmk:BCS}
\be{enumerate}[label=(\roman*)]
\item The formulation of the BCS model that we use is due to \cite{BLS94, deGennes}. A heuristic derivation from the quantum many body Hamiltonian can be found in the appendix to \cite{HHSS08}.
\item The matrix elements of a BCS state $\Gam$ have the following physical significance. If we write $\langle\cdot\rangle$ for the expectation value of an observable in the system state and $\gam(x,y),\al(x,y)$ for the operator kernels of $\gam,\al$, then $\gam(x,y)=\langle a_x^\dagger a_y\rangle$ is the one-particle density matrix and $\al(x,y)=\langle a_x a_y\rangle$ is the fermion pairing function. (Here $a_x^\dagger,a_x$ denote the fermion creation and annihilation operators as usual.) 

We will abuse notation and denote the kernel functions of $\gam$ and $\al$ by $\gam$ and $\al$ as well. 
\item We ignore spin variables. Implicitly, the pairing function $\al(x,y)$ (which is symmetric since $\al^*=\ol{\al}$) is to be tensored with a \emph{spin singlet}, yielding an antisymmetric two body wave function, as is required for fermions. 
\item For simplicity, we do not include an external magnetic field in the model. There is no apparent obstruction to applying the methods with a sufficiently regular and weak external magnetic field as in \cite{FHSS12, FHSS15, HS12}.
\e{enumerate}
\e{rmk}

Throughout, we make 

\be{ass}[Regularity of $V$ and $W$]
\label{ass:V}
$V:\R^d\to \R$ is a locally integrable function that is infinitesimally form-bounded with respect to $-\De$ (the ordinary Laplacian) and $V$ is reflection-symmetric, i.e.\ $V(x)=V(-x)$. Moreover, $-\De+V$ admits a ground state of negative energy $-E_b$. 

We also assume that $W\in L^{p_W}(\Om)$ with $2\leq p_W\leq \it$ if $d=1$, $2<p_W\leq \it$ if $d=2$ and $3\leq p_W\leq\it$ if $d=3$.
\e{ass}

\be{rmk}
\be{enumerate}[label=(\roman*)]
\item The assumption that $V$ admits a two-particle bound state is critical for the fermion pairs to condense. Without it, the pairs would prefer to drift far apart to be energy-minimizing. (Strictly speaking, each fermion pair is described by the operator $-2\De+2V$ and has the ground state energy $-2E_b$. We have made the factor two disappear for notational convenience, observe also the lack of a symmetrization factor $1/2$ in front of the $V$ term in \eqref{eq:EBCSdefn}.) 

\item The integrability assumption on $W$ is such that $W\psi\in L^2(\Om)$ for every $\psi\in H^1_0(\Om)$ and the numerical value of $p_W$ is derived from the critical Sobolev exponent. 

Note that the assumption implies that $W$ is infinitesimally form-bounded with respect to $-\Delta$. However, the assumption is stronger than infinitesimal form-boundedness (which would e.g.\ be guaranteed by $|W|^{1/2}\psi\in L^2(\Om)$ for every $\psi\in H^1_0(\Om)$) and the two places where we use this additional strength are (a) for the semiclassical expansion (Lemma \ref{lm:semiclassics}) and (b) for Davies' approximation result (Lemma \ref{lm:Davies}). 
\e{enumerate}
\e{rmk}


\be{ass}[Regularity of $\Om$]
\label{ass:Omega}
The open set $\Om\subseteq \R^d$ is a bounded Lipschitz domain.
\e{ass}

We recall that a set $\Om$ is a Lipschitz domain if its boundary can be locally represented as the graph of a Lipschitz continuous function. The formal definition is given in Appendix \ref{app:hardy}. 



\be{defn}[Admissible states]
We say that a BCS state $\Gam$ of the form \eqref{eq:Gamdefn} is \emph{admissible}, if $\Tr{\gam^{1/2}(1-\De_\Om)\gam^{1/2}}<\it$. 
\e{defn}

An admissible state $\Gam$ has the integral kernel $\al\in H_0^1(\Om^2)$ thanks to the operator inequality $\al\ol{\al}\leq \gam$ and $\al^*=\ol{\al}$ (we skip the proof, see the last step in the proof of Proposition \ref{prop:Gampsi} for a closely related argument).  We note

\be{prop}
\label{prop:boundedbelow}
$\curly{E}_\mu^{BCS}$ is bounded from below on the set of admissible states $\Gam$.
\e{prop}

In principle, this is a standard argument based on $\al\ol{\al}\leq \gam$ and our assumption that $V$ is infinitesimally form-bounded with respect to $-\De$. However, a little care has to be taken regarding the boundary conditions; we leave the proof to the interested reader because the required ideas appear throughout the paper. 

In this paper, we shall study the minimization problem 
\beq
\label{eq:EBCSmudefn}
E^{BCS}_\mu:=\inf_{\Gam \textnormal{ admissible}} \curly{E}_\mu^{BCS}(\Gam).
\eeq
 Note that $E^{BCS}_\mu>-\it$ by Proposition \ref{prop:boundedbelow}. We are especially interested in the occurrence of $E^{BCS}_\mu<0$ and in that case we say that the system exhibits \emph{fermion pairing}.

Here is the reasoning behind this definition: We will consider chemical potentials $\mu=-E_b+Dh^2$ with $D\in\R$ so that $\goth{h}\geq 0$ for $h$ small enough, see Proposition \ref{prop:gothh}. Then $E^{BCS}_\mu<0$ implies that any minimizer $\Gam$ must satisfy $\al\neq 0$, i.e.\ it must have a non-trival fermion pairing function $\al$. 


\paragraph{Main results.}
We now discuss our main results in words, they are stated precisely in Section \ref{ssect:mr} below.

By the monotonicity of $\mu\mapsto E_\mu^{BCS}$ for every fixed $h>0$, there exists a unique critical chemical potential $\mu_c(h)$ such that we have fermion pairing iff $\mu>\mu_c(h)$. The first natural question is then whether one can compute $\mu_c(h)$. In our \textbf{first main result}, \textbf{Theorem \ref{thm:main1}}, we show that 
$$
\mu_c(h)=-E_b+h^2 D_c + O(h^{2+\nu}), \quad \text{as } h\downarrow 0.
$$
That is, to lowest order in $h$, $\mu_c(h)$ is just one half of the binding energy of a fermion pair. The subleading correction term $D_c\in\R$ is the ground state energy of an explicit Dirichlet eigenvalue problem on $\Om$ (the linearization of the GP theory below).

Physically, the choice of $\mu\approx \mu_c(h)$ corresponds to small density; this is explained after Proposition \ref{prop:wc}. Therefore, we expect that for $\mu>\mu_c(h)$ the fermion pairs look like bosons to each other and (since we are at zero temperature) the pairs will form a Bose-Einstein condensate, which will then be describable by a Gross-Pitaevskii (GP) theory.

Accordingly, in our \textbf{second main result}, \textbf{Theorem \ref{thm:main2}}, we derive an effective, macroscopic GP theory of fermion pairs from the BCS model for all $\mu=-E_b+Dh^2$ with $D\in\R$. The resulting GP theory \emph{also has Dirichlet boundary conditions}. 

Theorems \ref{thm:main1} and \ref{thm:main2} show that the boundary conditions make a significant difference on the (macroscopic!) GP scale, a physically non-trivial fact. The results hold for the rather general class of bounded Lipschitz domains. 

\paragraph{Related works.} The BCS model that we consider has received considerable interest in recent years in mathematical physics. Most closely related to our paper are the derivations of effective GP theories for periodic boundary conditions in \cite{HS12} and for a system in $\R^3$ at fixed particle number \cite{B}. The time-dependent analogue of this derivation was performed in \cite{HS13}. The related, and technically more challenging, case of BCS theory close to the critical temperature for pair formation has also been considered: In \cite{FHNS07, HHSS08}, the critical temperature was described by a linear criterion. The analogue of Theorem \ref{thm:main1} for the upper and lower critical temperatures was the content of \cite{FHSS15}. In \cite{FHSS12c, FL15} and especially \cite{FHSS12} effective macroscopic Ginzburg-Landau theories have been derived. 

We emphasize that all of these papers assume that the system has no boundary (either by working on the torus or on the whole space) and the same holds true for the resulting effective GP or GL theories. (We also mention that the derivation in \cite{B} depends on $\|W\|_{L^\it(\R^d)}<\it$ and so one cannot obtain the Dirichlet boundary conditions as the limiting case of a sufficiently deep potential well from \cite{B}.)

Our main contribution is thus to show the \emph{non trivial effect of boundary conditions on the effective macroscopic GP theory}. As we mentioned in the introduction, this is in some contrast to de Gennes' arguments \cite{deGennes} at positive temperature.
%

 \subsection{Main result 1: The critical chemical potential}
Considering definitions \eqref{eq:EBCSdefn} and \eqref{eq:EBCSmudefn} of the BCS energy, we see that the non-positive function $\mu\mapsto E^{BCS}_\mu$ is monotone decreasing (and concave). This allows us to define the critical chemical potential $\mu_c(h)$ by
\beq
\label{eq:mucdefn}
\mu_c(h):=\inf\setof{\mu<0}{E^{BCS}_{\mu}<0}
\eeq
In other words, the definition is such that fermion pairing occurs iff $\mu>\mu_c(h)$. (Note that $\mu_c(h)$ may be infinite at this stage.) This is analogous to the definition of the upper and lower critical temperature in \cite{FHSS15}, but the explicit dependence of the BCS energy on $\mu$ simplifies matters here.

Our first main result gives an asymptotic expansion of $\mu_c(h)$ in $h$ up to second order, where the subleading term $D_c$ is given as an appropriate Dirichlet eigenvalue, namely
$$
D_c:=\inf\mathrm{spec}_{L^2(\Om)}\l(-\frac{1}{4}\De_\Om+W\r)
$$
The result is the analogue of the main result in \cite{FHSS15} for the critical temperature.
 
\be{thm}[Main result 1]
\label{thm:main1} 
We have 
$$
\mu_c(h)=-E_b+D_c h^2 +O(h^{2+\nu}), \quad \text{as } h \downarrow 0
$$ 
The exponent of the error term is $\nu:=\min\{d/2, c_\Om-\de\}$ where $\de>0$ is arbitrarily small and $c_\Om\in (0,1]$ depends only on $\Om$, see Remark \ref{rmk:main1} (iii) below.
\e{thm}
 
 \be{rmk}
 \label{rmk:main1}
\be{enumerate}[label=(\roman*)]
\item It follows from the definition of $D_c$ that the Dirichlet boundary conditions have a non-trivial effect on the value of $\mu_c(h)$. 

\item The critical value $D_c$ is uniquely determined by $E^{GP}_{D}=0$ for $D\leq D_c$ and $E^{GP}_{D}<0$ for $D>D_c$, where $E^{GP}_{D}$ is defined in \eqref{eq:EGPdefn} and \eqref{eq:EGPDdefn}. For the proof, see Lemma 2.5 in \cite{FHSS15}.
\item The constant $c_\Om$ in the definition of $\nu$ is the constant such that the Hardy inequality \eqref{eq:hardy} holds on $\Om$. Under additional assumptions on $\Om$, quantitative information on $c_\Om$ is known: If $\Om$ is convex or if $\del\Om$ is given as the graph of a $C^2$ function, then $c_\Om=1$ which is optimal \cite{BrezisMarcus, Marcusetal, MS} and if $\Om\subset\R^2$ is simply connected, then we can take $c_\Om=1/2$ \cite{Ancona86}. 
\item The asymptotic expansion of $\mu_c(h)$ to this order is \emph{the same as the expansion of the ground state energy of the two body Schr\"odinger operator} $H_h$, see Theorem \ref{thm:twobody}. Intuitively, this is due to the fact that at $\mu_c(h)$ fermion pairing just onsets, so the order parameter is small and the nonlinear terms become negligible.
\e{enumerate}
\e{rmk}

\subsection{Main result 2: Effective GP theory}
\label{ssect:mr}
\be{defn}
We write $\al_*$ for the unique positive and $L^2$-normalized ground state of $-\De+V$. By definition, it satisfies  $(-\De+V)\al_*=-E_b\al_*$. Let
\beq
\label{eq:gbcsdefn}
g_{BCS}:=(2\pi)^{-d}\int_{\R^d}(p^2+E_b)|\ft\al_*(p)|^4\d p.
\eeq
For any $D\in\R$ and $\psi\in H_0^1(\Om)$, we define the Gross-Pitaevskii (GP) energy functional by
\beq
\label{eq:EGPdefn}
\curly{E}^{GP}_{D}(\psi) :=\int_{\Om} \l(\frac{1}{4}|\nabla\psi(X)|^2+ (W(X)-D) |\psi(X)|^2+g_{BCS}|\psi(X)|^4\r) \d X.
\eeq
\e{defn}


We now state our second main result. It says that the GP theory $\curly{E}^{GP}_{D}$ arises from $\curly{E}_{-E_b+Dh^2}^{BCS}$ as the scale parameter $h$ goes to zero.

\be{thm}[Main result 2]
\label{thm:main2}
Let $\mu=-E_b+Dh^2$ for $D\in\R$ and define
\beq
\label{eq:EGPDdefn}
E_{D}^{GP}:=\inf_{\psi\in H_0^1(\Om)}\curly{E}_D^{GP}(\psi).
\eeq
\be{enumerate}[label=(\roman*)]
\item
As $h\downarrow0$,
\beq
\label{eq:main}
E^{BCS}_\mu=h^{4-d} E_D^{GP}+O(h^{4-d+\nu}),
\eeq
where $\nu$ is as in Theorem \ref{thm:main1}.
\item  Let $\Om$ be convex. Suppose that $\Gam$ is a BCS state such that
$$
\curly{E}_\mu^{BCS}(\Gam)\leq E_\mu^{BCS}+\eps h^{4-d}
$$
for some small $\eps>0$. Then, its upper right entry $\al$ in the sense of \eqref{eq:Gamdefn} can be decomposed as
\beq
\label{eq:aldecomp}
\al(x,y)=h^{1-d}\psi\l(\frac{x+y}{2}\r)\al_*\l(\frac{x-y}{h}\r)+\xi\l(\frac{x+y}{2},x-y\r)
\eeq
with $\psi\in H_0^1(\Om)$ satisfying $\curly{E}^{GP}_D(\psi)\leq E_D^{GP}+\eps+O(h^{\nu})$ and $\xi\in H_0^1(\Om\times\R^d)$
such that
$$
\|\xi\|_{L^2(\Om\times\R^d)}^2+h^2 \|\nabla\xi\|_{L^2(\Om\times\R^d)}^2\leq O(h^{4-d}).
$$
\e{enumerate}
\e{thm}

\be{rmk}
\label{rmk:main}
The interpretation of Theorem \ref{thm:main2} (ii) is that GP theory also describes the \emph{approximate minimizers} of BCS theory. 
If $\Om$ is not convex, one can still get a weaker version in which $\psi$ and the Dirichlet energy live on a slightly enlarged domain, see Theorem \ref{thm:key1} (LB).
\e{rmk}



We close the presentation by explaining why the choice of $\mu=-E_b+Dh^2$ corresponds to a low density limit.  

\be{prop}[Convergence of the one body density]
\label{prop:wc}
 Let $\Gam$ be a BCS state satisfying $\curly{E}_{-E_b+Dh^2}^{BCS}(\Gam)\leq E_{-E_b+Dh^2}^{BCS}+o(h^{4-d})$ (e.g.\ $\Gam$ is an approximate minimizer as in Theorem \ref{thm:main2} (ii)) and let $\rho_\gam$ denote its one body density matrix. Then we have
\beq
\label{eq:weakconv}
h^{d-2} \rho_\gam  \rightharpoonup |\psi_*|^2,\qquad \text{in } L^{p_W'}(\Om)
\eeq
where $\psi_*$ is a minimizer of $E^{GP}_D$ (minimizers exist and are unique up to a complex phase, see Proposition \ref{prop:existence}).  $p_W'$ is the H\"older dual of $p_W$.
\e{prop}

The proof is in Appendix \ref{app:wc}. This is a classical argument which is based on Theorem \ref{thm:main2} and the fact that the one body density $\rho_\gam$ and the external field $W$ are dual variables \cite{Griffiths,LiebSimonTF}.

Note that we can test \eqref{eq:weakconv} against $1_\Om$ to obtain the expected particle number 
$$
N:=\int_{\Om}\rho_\gam \d x=h^{2-d}\int_\Om |\psi_*|^2 \d x +o(h^{2-d}),
$$
compare (1.14) in \cite{HS12}. The expected particle density in microscopic units is given by $h^d N= h^2 \|\psi_*\|_{L^2(\Om)}^2+o(h^2)\to 0$. We see that our scaling limit indeed corresponds to low density. (We point out that the physical model is somewhat pathological in $d=1$ because even $N$ will go to zero as $h\to 0$. Since $N$ is only the \emph{expected} particle number, the model still makes sense in principle, but it is of course debatable that statistical mechanics still applies in this case.)

\subsection{Outline of the paper}
The proof of the main results is based on two distinct key results. 
\be{itemize}
\item In \textbf{key result 1} (Theorem \ref{thm:key1}), we bound the BCS energy over $\Om$ in terms of GP energies on a slightly smaller domain than $\Om$ (upper bound) and on a slightly larger domain than $\Om$ (lower bound). If $\Om$ is convex, the lower bound simplifies to the GP energy on $\Om$ itself. The general strategy here is as in \cite{FHSS12,HS13,HS12}, though some technical difficulties arise from the Dirichlet boundary conditions, see (i) and (ii) below. This part only requires $\Om$ to have finite Lebesgue measure.

\item In \textbf{key result 2} (Theorem \ref{thm:key2}), we show that the GP energy is \emph{continuous under approximations of the domain $\Om$}, if $\Om$ is a bounded Lipschitz domain. The idea is to use Hardy inequalities to control the boundary decay of GP minimizers using that these lie in the operator domain of the Dirichlet Laplacian. This approach is due to Davies \cite{Davies93, Davies00} who treated the linear case of Dirichlet eigenvalues. (Davies does not treat continuity under exterior approximations because a Hardy inequality is not sufficient for this to hold, see the example in Remark \ref{rmk:example})
\e{itemize}

We point out that key result 1 concerns the many body system. Key result 2, by contrast, is a continuity result for a certain class of nonlinear functionals on $\R^d$ and is based on ideas from spectral theory and geometry. 

In \textbf{Section \ref{sect:key}}, we present the two key results and derive the two main results from them. 

In \textbf{Section \ref{sect:semiclassics}}, we present the semiclassical expansion (Lemma \ref{lm:semiclassics}). This is an important tool in the proof of all parts of Theorem \ref{thm:key1} (key result 1). The version here is very close to the one in \cite{B}, though we generalize it somewhat as described in (iii) below. (In \cite{B}, an idea from \cite{HS13} was used to simplify the semiclassics significantly in the zero temperature case as compared to \cite{FHSS12, HS12}.)

In \textbf{Section \ref{sect:UB}}, we prove the upper bound part of Theorem \ref{thm:key1}. We construct a trial state following \cite{B,HS13}, with an appropriate cutoff to ensure that it satisfies the Dirichlet boundary conditions. The semiclassical expansion then yields an upper bound by a GP energy in a slightly smaller region than $\Om$. One finishes the proof by applying the continuity of the GP energy under domain approximations (key result 2).

In \textbf{Sections \ref{sect:LBA}-\ref{sect:LBB}}, we prove the lower bound part of Theorem \ref{thm:key1}. The overall strategy is as in \cite{B,FHSS12}: One first proves an a priori decomposition result yielding \eqref{eq:aldecomp} for the off diagonal entry $\al$ of any approximate BCS minimizer $\Gam$ (with $H^1$ control on the involved functions). This is Theorem  \ref{thm:apriori} and it shows that the GP order parameter is naturally associated with the center of mass variable $\frac{x+y}{2}$ (living on the macroscopic scale). Then, one can use the semiclassical expansion on the main part of $\al$ to finish the proof.

While the overall strategy is as in \cite{B,FHSS12}, there are some additional technical difficulties, mainly due to the boundary conditions:

\be{enumerate}[label=(\roman*)]
\item The boundary conditions prevent the variables in the center of mass frame from decoupling as usual. This poses a problem, because the GP energy/order parameter should only depend on the center of mass variable. The solution we have found to this is to \emph{forget the boundary conditions in the relative coordinate altogether}. (Note that this gives a lower bound, since Dirichlet energies decrease under an increase of the underlying function spaces.) In this way, we \emph{decouple the variables} in the center of mass frame. Moreover, one has not lost much, thanks to the exponential decay of the Schr\"odinger eigenfunction $\al_*$ governing the relative coordinate via \eqref{eq:aldecomp}.
\item The center of mass variable $\frac{x+y}{2}$ naturally takes values in the set 
$$
\tilde\Om:=\frac{\Om+\Om}{2}.
$$
After some steps in the lower bound, we are led to a GP energy on $\tilde\Om$. Note that when $\Om$ is convex, $\tilde\Om=\Om$ and so one is essentially done at this stage. If $\Om$ is not convex, however, some additional work is required. The idea is to use the exponential decay of $\al_*$ again, the details are in Section \ref{sect:LBnonC}.
\item We observe that the arguments from \cite{B} can be extended to dimensions $d=1,2$ and to external potentials which satisfy $W\in L^{p_W}(\Om)$. We do not see, however, that the arguments can be extended to the case $W=\it$ on a set of positive measure (i.e.\ the Dirichlet boundary conditions). 
\e{enumerate}

In \textbf{Section \ref{sect:key2}}, we prove key result 2, Theorem \ref{thm:key2}. The crucial input are Davies' ideas \cite{Davies93, Davies00} of deriving continuity of the Dirichlet energy under domain approximations from the Hardy inequality, see Lemma \ref{lm:Davies}. Along the way, we need Theorem \ref{thm:hardy} which says that the Hardy inequality holds along a suitable sequence of exterior approximations $\Om_\ell$ to $\Om$, with uniform dependence of the Hardy constants on $\ell$, and may be of independent interest.

Theorem \ref{thm:hardy} is proved in Appendix \ref{app:hardy} by extending Necas' proof \cite{Necas} of the Hardy inequality on any bounded Lipschitz domain. The appendix also contains the proofs of some technical results used in the main text, as well as a presentation of the \emph{linear version of our main results}, the asymptotics of the ground state energy of the two body Schr\"odinger operator $H_h$ mentioned in the introduction (see Appendix \ref{app:linear}).

\be{rmk}[Notation]
We write $C,C',\ldots$ for positive, finite constants whose value may change from line to line. We typically do not track their dependence on parameters which are assumed to be fixed throughout, such as the dimension $d$ and the potentials $V$ and $W$. The dependence on $D$ will be explicit only where relevant.

We will denote $\curly{E}^{BCS}_\mu\equiv \curly{E}^{BCS}, \curly{E}^{GP}_D\equiv \curly{E}^{GP},$ etc.
\e{rmk}

\section{The two key results}
\label{sect:key}
\subsection{Key result 1: Bounds on the BCS energy}
We bound the BCS energy on $\Om$ in terms of GP energies on interior approximations of $\Om$ for an upper bound (``UB'') and on exterior approximations of $\Om$ for a lower bound (``LB''). To state the result, we need to define the GP energy on a general domain $U\subseteq \R^d$.

\be{defn}[GP energy on domains]
For a finite-measure domain $U\subseteq \R^d$ and any $\psi\in H_0^1(U)$, we define the GP energy by
\beq
\label{eq:EGPUdefn}
\curly{E}^{GP}_{U}(\psi) :=\int_{U} \l(\frac{1}{4}|\nabla\psi(X)|^2+ (W(X)-D) |\psi(X)|^2+g_{BCS}|\psi(X)|^4\r) \d X,
\eeq
with $g_{BCS}$ as in \eqref{eq:gbcsdefn}. Here we extended $W:\Om\to \R$ by zero to get a map on $\R^d$. 
\e{defn}

Note that $\curly{E}^{GP}_{U}$ implicitly depends on $D$. We can now state

\be{thm}[Key result 1]
\label{thm:key1}
Let $\Om\subset\R^d$ be an open set of finite Lebesgue measure. For $\ell>0$, define the interior and exterior approximations of $\Om$
\begin{align}
\label{eq:Omell-defn}
\Om_{\ell}^-:=&\setof{X\in \Om}{\mathrm{dist}(X,\Om^c)> \ell},\\
\label{eq:Omell+defn}
\Om_{\ell}^+:=&\setof{X\in \R^d}{\mathrm{dist}(X,\Om)< \ell}.
\end{align}
Let $\ell(h):=h\log(h^{-q})$ with $q>0$ sufficiently large but fixed. Then:
\be{enumerate}
\item[(UB)]
For every function $\psi\in H_0^1(\Om_{\ell(h)}^-)$, there exists an admissible BCS state $\Gam_\psi$ such that
\beq
\label{eq:UB}
\curly{E}^{BCS}(\Gam_\psi)= h^{4-d}\curly{E}^{GP}_{\Om_{\ell(h)}^-}(\psi)+O(h^{5-d})(\|\psi\|_{H^1(\R^d)}^2+\|\psi\|_{H^1(\R^d)}^4).
\eeq
with $D=h^{-2}(\mu+E_b)$.
\item[(LB)]
Let $\mu\leq -E_b+O(h^2)$ and let $\Gam$ be an admissible BCS state satisfying $\curly{E}^{BCS}(\Gam)\leq C_\Gam h^{4-d}$. Then, there exist $\psi\in H_0^1(\Om_{\ell(h)}^+)$ such that 
\beq
\label{eq:LB}
\curly{E}^{BCS}(\Gam)
\geq h^{4-d} \curly{E}^{GP}_{\Om_{\ell(h)}^+}(\psi) + O(h^{4-d+\nu'}),
\eeq
where $\nu'=\min\{d/2,1\}$. Moreover, there exists $\xi\in H_0^1(\tilde\Om\times\R^d)$, $\tilde\Om:=\frac{\Om+\Om}{2}$ such that $\al$ can be decomposed as in \eqref{eq:aldecomp} and we have the bounds
\beq
\label{eq:keybounds}
\begin{aligned}
&\|\nabla\psi\|_{L^2(\Om_{\ell(h)}^+)}\leq C\|\psi\|_{L^2(\Om_{\ell(h)}^+)}\leq O(1),\\
&\|\xi\|_{L^2(\tilde\Om\times\R^d)}^2+h^2 \|\nabla\xi\|_{L^2(\tilde\Om\times\R^d)}^2\leq O(h^{4-d}) (\|\psi\|_{L^2(\tilde\Om)}^2+C_\Gam)
\end{aligned}
\eeq

\item[(LBC)] If $\Om$ is convex, then one can take $\ell(h)=0$ everywhere in (LB). In particular, there exists $\psi\in H_0^1(\Om)$ such that
\beq
\label{eq:LBC}
\curly{E}^{BCS}(\Gam)
\geq h^{4-d} \curly{E}^{GP}_\Om(\psi) + O(h^{4-d+\nu'}).
\eeq

\e{enumerate}
\e{thm}

\subsection{Key result 2: Continuity of the GP energy under domain approximations}
\label{sect:Davies}

On any bounded Lipschitz domain, we have continuity of the GP energy under domain approximations. The continuity is derived from the Hardy inequality \eqref{eq:hardy} in an approach due to Davies \cite{Davies93,Davies00}, see also \cite{Evansetal}. The details are in Section \ref{sect:key2}. Define
$$
E^{GP}_U:=\min_{\psi\in H_0^1(U)} \curly{E}^{GP}(U).
$$

\be{thm}
\label{thm:key2}
Assume that $\Om$ is a bounded Lipschitz domain. For $\ell>0$, define $\Om_\ell^\pm$ as in Theorem \ref{thm:key1}. Then, there exists a constant $c_\Om\in (0,1]$ such that
\beq
\label{eq:continuity}
|E^{GP}_{\Om_{\ell}^\pm}-E^{GP}_\Om|\leq O(\ell^{c_\Om}).
\eeq
Moreover, the statement holds irrespectively of the value of the parameters $g_{BCS}$ and $D$ in \eqref{eq:EGPUdefn}. In particular it holds for $g_{BCS}=D=0$ and then it shows that
\beq
\label{eq:Dcelldefn}
|D_c^\pm(\ell)-D_c|\leq O(\ell^{c_\Om}),\qquad 
D_c^\pm(\ell):=\inf\mathrm{spec}_{L^2(\Om)}\l(-\frac{1}{4}\De_{\Om_{\ell}^\pm}+W\r).
\eeq
\e{thm}

\be{rmk}
\be{enumerate}[label=(\roman*)]
\item
The constant $c_\Om$ is the same as in Theorem \ref{thm:main1}; see Remark \ref{rmk:main} for quantitative results on $c_\Om$ if more information on $\Om$ is known.

\item
The implicit constant in \eqref{eq:continuity} depends on $D$ and consequently the same is true for the implicit constant in \eqref{eq:main}. It will be important in the proof of Theorem \ref{thm:main1} that the $D$-dependence has disappeared in \eqref{eq:Dcelldefn}.
\e{enumerate}
\e{rmk}



We close with a cautionary example, which shows that the assumption that $\Om$ is a Lipschitz domain is rather sharp for getting a two-sided continuity result such as \eqref{eq:continuity}.

\be{rmk}[Exterior approximation is delicate]
\label{rmk:example}
Consider the slit domain $\Om=[-1,1]^2\setminus ((-1,0]\times\{0\})$. The slit will disappear for any sequence of exterior approximations and this will lead to an order one decrease of the GP energy. Therefore, the GP energy on $\Om$ is \emph{not continuous under exterior approximation}. (However, it is continuous under interior approximation: As discussed in Section \ref{sect:hardy}, this follows from the validity of the Hardy inequality \eqref{eq:hardy} on $\Om$, and since $\Om\subset \R^2$ is simply connected, it satisfies the Hardy inequality with  $c_\Om=1/2$ \cite{Ancona86}.)
\e{rmk}

\subsection{On GP minimizers}

We collect some standard results about GP minimizers for later use: They exist (though they may be identically zero) and their $-\Delta_U$ operator norm is bounded by their $H^1_0$ norm (because they satisfy the Euler-Lagrange equation). 

\be{prop}
\label{prop:existence}
Let $U\subset\R^d$ be an open set of finite Lebesgue measure. 
\be{enumerate}[label=(\roman*)]
\item For any $\psi\in H_0^1(U)$, we have the coercivity 
\beq
\label{eq:coercivity}
\curly{E}^{GP}_U(\psi)\geq C_1\|\psi\|_{H_0^1(U)}^2-(C_2+D)^2,
\eeq
where the constants $C_1,C_2>0$ are independent of $U$ and $D$. In particular, $E_U^{GP}:=\inf_{\psi\in H_0^1(U)}\curly{E}^{GP}_U(\psi)>-\it$.

\item There exists a minimizer for $\curly{E}^{GP}_U$ and it is unique up to multiplication by a complex phase. Moreover, minimizing sequences are precompact in $H_0^1(U)$.

\item There exists $C>0$, independent of $U$ and $D$, such that the minimizer $\psi_*$ satisfies 
\beq
\label{eq:Debound}
\|\Delta_U\psi_*\|_{L^2(U)}\leq C(1+|D|) (\|\psi_*\|_{H^1_0(U)}+\|\psi_*\|_{H^1_0(U)}^3).
\eeq
\e{enumerate}
\e{prop}

For completeness, the standard proof of these results is included in Appendix \ref{app:existence}. 

\subsection{Proof of the main results from the key results}
\label{sect:proofmain}

In this section, we assume that the two key results (Theorems \ref{thm:key1} and \ref{thm:key2}) hold. 

\subsubsection{Proof of main result 1, Theorem \ref{thm:main1}}
\paragraph{Upper bound.}
We will show that there exists a constant $C_0>0$ such that for all $\mu= -E_b+Dh^2$ with $D\geq D_c+C_0h^\nu$, there exists an admissible BCS state $\Gam$ such that 
\beq
\label{eq:muUB}
\curly{E}^{BCS}(\Gam)<0.
\eeq
By Definition \eqref{eq:mucdefn}, this implies the claim $\mu_c(h)\leq -E_b+D_ch^2+C_0h^{2+\nu}$.

We let $\ell\equiv \ell(h)=h\log(h^{-q})$ with $q>0$ large enough and we recall definitions \eqref{eq:Omell-defn} and \eqref{eq:Dcelldefn} of $\Om_\ell^-$ and $D_c^-(\ell)$.
Following \cite{FHSS15} p.209, we choose $\psi=\theta\psi_{\ell}$, where $\theta>0$ and $\psi_{\ell}\in H_0^1(\Om^-_\ell)$ is the eigenfunction 
$$
(-\De_{\Om_\ell^-}+W)\psi_{\ell}=D_c^-(\ell)\psi_\ell.
$$
Our Assumption \ref{ass:V} on $W$ implies that $\psi_\ell\in\mathrm{dom}(\De_{\Om_\ell^-})$. Optimizing over $\theta$ yields 
\beq
\label{eq:DDC}
\curly{E}_{\Om_\ell^-}^{GP}(\psi)=-C(D-D_c^-(\ell))^2,\qquad \theta=C'\sqrt{D-D_c^-(\ell)}.
\eeq
Hence, any relevant norm of $\psi=\theta\psi_\ell$ is proportional to $\sqrt{D-D_c^-(\ell)}$. Since $\psi\in H_0^1(\Om_{\ell(h)}^-)$, we can apply Theorem \ref{thm:key1} (UB) to get an admissible BCS state $\Gam_{\psi}$ such that
$$
\begin{aligned}
h^{d-4}\curly{E}^{BCS}(\Gam_{\psi})
=& \curly{E}^{GP}_{\Om_{\ell}^-}(\psi)+O(h^\nu)(\|\psi\|_{H^1(\R^d)}^2+\|\psi\|_{H^1(\R^d)}^4)\\
=& -C(D-D^-_c(\ell))^2+O(h^\nu)(\theta^2\|\psi_\ell\|_{H^1(\R^d)}^2+\theta^4\|\psi_\ell\|_{H^1(\R^d)}^4).
\end{aligned}
$$
We have the a priori bound $\|\psi_\ell\|_{H^1(\R^d)}\leq O(1)$. Indeed, the infinitesimal-form boundedness of $W$ with respect to $-\De_{\Om_\ell^-}$ implies
$$
\|\psi_\ell\|_{H^1(\R^d)}-C\leq D_c^-(\ell)\leq D_c^-(\ell_0),
$$
where $\ell_0>0$ is fixed. In the second step, we used the fact that Dirichlet energies increase when the underlying domain decreases.

By our choice of $D$ and the last part of Theorem \ref{thm:key2}, there exists $C_1>0$ such that
$$
D>D_c+C_0h^\nu\geq D_c^-(\ell)+(C_0-C_1)h^\nu
$$
and so, for $C_0>C_1$,
$$
h^{d-4}\curly{E}^{BCS}(\Gam_{\psi})=-C(C_0-C_1)^2 h^{2\nu} +O(h^{2\nu})(C_0-C_1).
$$
Clearly, this can be made negative by choosing $C_0>0$ large enough. This proves \eqref{eq:muUB} and hence the claimed upper bound on $\mu_c(h)$.
\qed

\paragraph{Lower bound (convex case).}
 Let $\mu=-E_b+Dh^2$ with $D<D_c-C_0h^{\nu}$ with $C_0$ to be determined. Let $\Gam$ be a BCS state satisfying $\curly{E}^{BCS}(\Gam)\leq 0$. We will show that $\Gam\equiv 0$ and this will prove the claim $\mu_c(h)\geq -E_b +h^2 D_c-C_0 h^{2+\nu}$.
 
 Assumption \ref{ass:V} on $W$ implies that it is infinitesimally form-bounded with respect to $-\De_\Om$ on $H_0^1(\Om)$ and from this one derives that $\goth{h}\geq 0$ for sufficiently small $h$, see Proposition \ref{prop:gothh}. Therefore, the zero state is the unique minimizer of the first term $\tr{\goth{h}\gam}$ in $\curly{E}^{BCS}$ and it suffices to show that $\al\equiv 0$ to get $\Gam=0$.

We apply Theorem \ref{thm:key1} (LBC) with $C_\Gam=0$ and obtain $\psi\in H_0^1(\Om)$ such that
$$
0\geq h^{d-4}\curly{E}^{BCS}(\Gam)
\geq  \curly{E}^{GP}_{\Om}(\psi) + O(h^{\nu})\|\psi\|_{H^1_0(\Om)}^2.
$$
We drop the (non-negative) quartic term in $\curly{E}^{GP}_{\Om}$ for a lower bound and use the definition of $D_c$ to get 
$$
\curly{E}^{GP}_{\Om}(\psi)\geq (D_c-D) \|\psi\|_{L^2(\Om)}^2
$$
The analogue of the first relation in \eqref{eq:keybounds} in the convex case is $\|\psi\|_{H^1_0(\Om)}^2\leq C\|\psi\|_{L^2(\Om)}^2$. It gives
\beq
\label{eq:onlydifference}
0\geq (C(D_c-D) +O(h^{\nu}))\|\psi\|_{H^1_0(\Om)}^2.
\eeq
Recall that $D_c-D>C_0h^{\nu}$. For $C_0$ large enough, this implies that $\psi\equiv 0$. Since $C_\Gam=0$, the analogue of the second bound in \eqref{eq:keybounds} in the convex case yields $\xi\equiv 0$ and so  $\al\equiv 0$ as claimed.

\paragraph{Lower bound (non convex case).}
We write $\ell\equiv \ell(h)$ throughout.
We apply Theorem \ref{thm:key1} (LB) and argue as in the convex case to find
$$
0\geq h^{d-4}\curly{E}^{BCS}(\Gam)
\geq   (D_c^+(\ell)-D+O(h^\nu))\|\psi\|_{L^2(\Om)}^2.
$$
Now, the last part of Theorem \ref{thm:key2} gives $D_c^+(\ell)-D+O(h^\nu)=D_c-D+O(h^\nu)$. This can be made positive by choosing $C_0$ large enough, which implies $\psi=0$ and so $\xi=0$ by \eqref{eq:keybounds}. This completes the proof of Theorem \ref{thm:main1}.
\qed

\subsubsection{Proof of main result 2, Theorem \ref{thm:main2}}
\label{sect:proofmain1}

We let $\mu=-E_b+Dh^2$ with $D\in \R$ fixed and we let $\ell(h)=h\log(h^{-q})$, with $q\geq 1$ large but fixed. 

\paragraph{Upper bound.}
By Proposition \ref{prop:existence}, the GP energy $\curly{E}^{GP}_{\Om_{\ell(h)}^-}$ has a unique minimizer, call it $\psi_-\in H_0^1(\Om_{\ell(h)}^-)$. We apply Theorem \ref{thm:key1} (UB) with $\psi=\psi_-$ to obtain an admissible BCS state $\Gam_{\psi_-}$ such that
$$
\begin{aligned}
E^{BCS}\leq \curly{E}^{BCS}(\Gam_{\psi_-})
=& h^{4-d}\curly{E}^{GP}_{\Om_{\ell(h)}^-}(\psi_-)+O(h^{5-d})(\|{\psi_-}\|_{H^1(\R^d)}^2+\|{\psi_-}\|_{H^1(\R^d)}^4)\\
\leq& h^{4-d}E^{GP}_{\Om_{\ell(h)}^-}+O(h^{5-d})(1+E^{GP}_{\Om_{\ell(h)}^-})^2.
\end{aligned}
$$
In the second step, we used the fact that $\psi_-$ is a minimizer and the coercivity \eqref{eq:coercivity}. 

Now we apply Theorem \ref{thm:key2}. Since $\ell(h)=O(h^{1-\de})$ for every $\de>0$, we get 
$$
E^{BCS}\leq  h^{4-d}E^{GP}_\Om+O(h^{4-d+\nu}),
$$
where $\nu$ is as in Theorem \ref{thm:main1}. 

\paragraph{Lower bound.}
Thanks to the upper bound right above, there exists $C>0$ such that for all $\eps>0$ we can find an approximate minimizer $\Gam$ such that
$$
 \curly{E}^{BCS}(\Gam)\leq h^{4-d}(E^{GP}_\Om+\eps C).
$$
In particular, $\curly{E}^{BCS}(\Gam)\leq C_\Gam h^{4-d}$ and so $\Gam$ satisfies the assumption in Theorem \ref{thm:key1} (LB) and (LBC).

If $\Om$ is convex, the claim follows directly from Theorem \ref{thm:key1} (LBC).

If $\Om$ is a non convex bounded Lipschitz domain, Theorem \ref{thm:key1} (LB) yields $\psi\in H_0^1(\Om_{\ell(h)}^+)$ such that
$$
\curly{E}^{BCS}(\Gam)
\geq h^{4-d} \curly{E}^{GP}_{\Om_{\ell(h)}^+}(\psi) + O(h^{4-d+\nu'})
\geq h^{4-d} E^{GP}_{\Om_{\ell(h)}^+} + O(h^{4-d+\nu'}).
$$
The lower bound now follows from Theorem \ref{thm:key2}. This finishes the proof of Theorem \ref{thm:main2}.
\qed

\section{Semiclassical expansion}
\label{sect:semiclassics}
We state an important tool for the proof of Theorem \ref{thm:key1}, the semiclassical expansion. The version here is essentially the one from \cite{B}.

Though not strictly necessary for the result, it will be convenient for us to assume the following decay condition

\be{defn}
\label{defn:expdecay}
We say that a function $\goth{a}\in L^2(\R^d)$ decays exponentially in the $L^2$ sense with the rate $\rho$, if
\beq
\label{eq:assexpdecay}
\int_{\R^d} e^{2\rho |s|}|\goth{a}(s)|^2 \d s<\it. 
\eeq
\e{defn}

Recall that $\al_*$ denotes the unique ground state of $-\De+V$. It is well known that weak assumptions on the potential $V$ imply the exponential decay of $\al_*$ in an $L^2$ sense. The fact that infinitesimal form-boundedness of $V$ is sufficient is essentially contained in \cite{Simon} but was known to the experts even earlier. 
That is, there exists $\rho_*>0$ such that
\beq
\label{eq:alexpdecay}
\int_{\R^d} e^{2\rho_* |s|}|\al_*(s)|^2 \d s<\it. 
\eeq
In particular, we can apply the following lemma with $\goth{a}=\al_*$ later on.

\be{lm}[Semiclassics]
\label{lm:semiclassics}
For $\psi, \goth{a}\in H^1(\R^d)$, we set
\beq
\label{eq:gothapsidefn}
\goth{a}_\psi(x,y):=h^{-d}\psi\l(\frac{x+y}{2}\r)\goth{a}\l(\frac{x-y}{h}\r),\quad x,y\in\R^d.
\eeq
Suppose that $\goth{a}(x)=\goth{a}(-x)$ and that $\goth{a}$ decays exponentially in the $L^2$ sense of Definition \ref{defn:expdecay}.

Then:
\be{enumerate}[label=(\roman*)]
\item
$$
\begin{aligned}
&\Tr{(-h^2\Delta-\mu)\goth{a}_\psi\ol{\goth{a}_\psi}}+\iint\limits_{\R^d\times\R^d} V\l(\frac{x-y}{h}\r) |\goth{a}_\psi(x,y)|^2\d x \d y\\
= &h^{-d}\|\psi\|^2_{L^2(\R^d)} \qmscp{\goth{a}}{-\De+E_b+V}{\goth{a}}\\
&+ \|\goth{a}\|_{L^2(\R^d)}^2\l(\frac{h^{2-d}}{4}\|\nabla\psi\|^2_{L^2(\R^d)}+h^{-d}(-E_b-\mu) \|\psi\|^2_{L^2(\R^d)}\r) .
\end{aligned}
$$
\item There exists a constant $C>0$ such that
$$
\begin{aligned}
&\l|\Tr{W\goth{a}_\psi\ol{\goth{a}_\psi}} -h^{-d} \|\goth{a}\|_{L^2(\R^d)}^2\int_{\R^d} W(X) |\psi(X)|^2 \d X\r|\\
\leq &C h^{1-d}  \|\goth{a}\|_{L^2(\R^d)}^2\|W\|_{L^{p_W}(\Om)}\|\psi\|_{H^1(\R^d)}^2.
\end{aligned}
$$

\item Let 
\beq
\label{eq:gbcsgothadefn}
\begin{aligned}
g_{BCS}(\goth{a}):&=(2\pi)^{-d}\int_{\R^d}(p^2+E_b)|\hat{\goth{a}}(p)|^4\d p,\\
g_0(\goth{a}):&=(2\pi)^{-d}\int_{\R^d}|\hat{\goth{a}}(p)|^4\d p
\end{aligned}
\eeq
Then, as $h\downarrow 0$,
$$
\begin{aligned}
\Tr{(-h^2\Delta+E_b+h^2W)\goth{a}_\psi\ol{\goth{a}_\psi}\goth{a}_\psi\ol{\goth{a}_\psi}}
&= h^{-d} g_{BCS}(\goth{a}) \|\psi\|_{L^4(\R^d)}^4+O(h^{1-d})\|\psi\|_{H^1(\R^d)}^4,\\
\Tr{\goth{a}_\psi\ol{\goth{a}_\psi}\goth{a}_\psi\ol{\goth{a}_\psi}}
&= h^{-d} g_{0}(\goth{a}) \|\psi\|_{L^4(\R^d)}^4+O(h^{1-d})\|\psi\|_{H^1(\R^d)}^4
\end{aligned}
$$
\e{enumerate}
\e{lm}

Lemma \ref{lm:semiclassics} was proved in in \cite{B} for $d=3$, $\goth{a}=h\al_*$, $W\in L^\it(\R^3)$ and at fixed particle number. We sketch the proof in Appendix \ref{app:semiclassics} to show that it generalizes to the present version.

\be{rmk}
\label{rmk:semi}
\be{enumerate}[label=(\roman*)]
\item We can apply the expansion in our situation because we can isometrically embed $H_0^1(U)\subset H^1(\R^d)$ by extending functions by zero. 

\item To see that $g_{BCS}(\goth{a}),g_{0}(\goth{a})<\it$, observe that the decay assumption \eqref{eq:assexpdecay} implies $\goth{a}\in L^1(\R^d)\cap H^1(\R^d)$ and so $\ft{\goth{a}}$ is bounded. 
\e{enumerate}
\e{rmk}

\section{Proof of Theorem \ref{thm:key1} (UB)}
\label{sect:UB}
The idea of the proof is to construct an appropriate trial state and then to use the semiclassical expansion from Lemma \ref{lm:semiclassics}.

\subsection{The trial state}
The trial state $\Gam_\psi$ is defined as in \cite{B}, following an idea of \cite{HS13}, see \eqref{eq:Gampsidefn} below. However, we multiply $\al_*$ by an appropriate cutoff function $\chi$, in order to satisfy the Dirichlet boundary conditions in the relative variable. 

\be{defn}[Trial state]
\label{defn:trial}
Let $\chi\in C_c^\it(\R^d)$ be a symmetric cutoff function, i.e. $\chi(r)=\chi(-r)$, $0\leq \chi\leq 1$ and $\chi\equiv 1$ on $B_1$ and $\mathrm{supp}\chi\subset B_{3/2}$. Let $\ell(h)=h\phi(h)$ with $\lim_{h\to 0}\phi(h)=\it$ and define
\beq
\label{eq:gothadefn}
\goth{a}(r):=\chi\l(\frac{r}{\phi(h)}\r) h \al_*(r).
\eeq
For any $\psi\in H^1(\R^d)$, we define $\goth{a}_\psi$ by \eqref{eq:gothapsidefn} and
\beq
\label{eq:Gampsidefn}
\gam_\psi:=\goth{a}_\psi\ol{\goth{a}_\psi}+(1+h^{1/2})\goth{a}_\psi\ol{\goth{a}_\psi}\goth{a}_\psi\ol{\goth{a}_\psi},\qquad \Gam_\psi:=\ttmatrix{\gam_\psi}{\goth{a}_\psi}{\ol{\goth{a}_\psi}}{1-\ol{\gam_\psi}}.
\eeq
\e{defn}

\be{prop}
\label{prop:Gampsi}
Let $\psi\in H_0^1(\Om)$.
For all sufficiently small $h$, $\Gam_\psi$ is an admissible BCS state.
\e{prop}

\be{proof}
$0\leq \Gam_\psi\leq 1$ holds by a short computation, see \cite{B}. We show that $\goth{a}_\psi\in H_0^1(\Om^2)$.  
First, we observe that $\supp \goth{a}_\psi\subseteq \Om^2$. To see this, we note that $\supp \psi\subseteq \Om_{\ell(h)}^{-}$ and $\supp \goth{a}\subseteq \supp \chi(\cdot/\phi(h))\subseteq B_{3\phi(h)/2}$ and therefore
$$
\supp \goth{a}_\psi\subseteq\setof{(x,y)\in\R^d\times \R^d}{\frac{x+y}{2}\in\Om_{\ell(h)}^{-},\;
 \frac{x-y}{2}\in B_{3\ell(h)/4}},
$$
where we also used $h\phi(h)=\ell(h)$. By construction, $\mathrm{dist}(\frac{x+y}{2},\Om^c)\geq \ell(h)$ and by expressing
$$
(x,y)=\l(\frac{x+y}{2}+\frac{x-y}{2},\frac{x+y}{2}-\frac{x-y}{2}\r),
$$
we obtain that, indeed, $\supp \goth{a}_\psi\subseteq \Om^2$.

It remains to show that, after extending $\psi$ and $\goth{a}$ by zero to $\R^d$, we have $\goth{a}_\psi\in H^1(\R^d\times \R^d)$. By using $\goth{a}(r)=\goth{a}(-r)$ to symmetrize the derivatives and changing to center-of-mass coordinates \eqref{eq:com}, we indeed get an upper bound on $\|\goth{a}_\psi\|_{H^1(\R^d\times \R^d)}$ in terms of the (finite) quantities $\|\psi\|_{H^1(\R^d)}$ and $\|\goth{a}\|_{H^1(\R^d)}$. We leave the details to the reader, as similar computations appear several times in the lower bound, see e.g.\ the proof of Lemma \ref{lm:apriori1}.

This proves $\goth{a}_\psi\in H_0^1(\Om^2)$. To see that $\gam$ is an $H_0^1$-density matrix, we note that $\gam_\psi\leq 3\goth{a}_\psi\ol{\goth{a}_\psi}$ since $\ol{\goth{a}_\psi}\goth{a}_\psi\leq \ol{\gam_\psi}\leq 1$. We can then bound
$$
\sqrt{1-\De_\Om}\gam_\psi\sqrt{1-\De_\Om}\leq 3\sqrt{1-\De_\Om}\goth{a}_\psi\ol{\goth{a}_\psi}\sqrt{1-\De_\Om}=3\sqrt{1-\De_\Om}\goth{a}_\psi\l(\sqrt{1-\De_\Om}\goth{a}_\psi\r)^*
$$
by a product of two Hilbert Schmidt operators and therefore it is trace class.
\e{proof}

\subsection{Controlling the effect of the cutoff}
When we apply the semiclassical expansion in Lemma \ref{lm:semiclassics}, we want to remove the effect of the cutoff, i.e. we want to replace $\goth{a}$ by $\al_*$, up to higher order corrections. We will get this from the estimates in Proposition \ref{prop:chi} below, which follow essentially from the exponential decay \eqref{eq:alexpdecay} of $\al_*$.

We recall definition \eqref{eq:gbcsgothadefn} of $g_{BCS}(\goth{a})$ and $g_0(\goth{a})$.

\be{prop}
\label{prop:chi}

Then, 
\begin{align}
\label{eq:chi1}
&\|\goth{a}\|_{L^2(\R^d)}^2=h^2\l(\|\al_*\|_{L^2(\R^d)}^2+O(e^{-2\rho_* \phi(h)})\r)\equiv h^2\l(1+O(e^{-2\rho_* \phi(h)})\r),\\
\label{eq:chi2}
&g_{BCS}(\goth{a})=h^4\l(g_{BCS}(\al_*)+O(e^{-\rho_* \phi(h)})\r)\equiv h^4\l(g_{BCS}+O(e^{-\rho_* \phi(h)/2})\r)\\
\label{eq:chi4}
&g_{0}(\goth{a})=h^4\l(g_{0}(\al_*)+O(e^{-\rho_* \phi(h)/2})\r)\\
\label{eq:chi3}
&\qmscp{\goth{a}}{-\De+E_b+V}{\goth{a}}=h^2 O(e^{-2\rho_* \phi(h)})
\end{align}
\e{prop}


\be{proof}
For \eqref{eq:chi1}, we observe
$$\begin{aligned}
\|h\al_*\|_{L^2(\R^d)}^2-\|\goth{a}\|_{L^2(\R^d)}^2
=&h^2 \int_{\R^d} |\al_*(r)|^2 \l(1-\chi\l(\frac{r}{\phi(h)}\r)^2\r) \d r\\
\leq &h^2\int_{B_{\phi(h)}^c} |\al_*(r)|^2 \d r \leq C h^2 e^{-2\rho_* \phi(h)}.
\end{aligned}
$$
In the last step, we used the fact that $\al_*$ satisfies the decay assumption \eqref{eq:alexpdecay}.

To get \eqref{eq:chi2}, we first write
\beq
\label{eq:from}
|h\ft\al_*|^4-|\hat{\goth{a}}|^4=\l(|h\ft\al_*|^2+|\hat{\goth{a}}|^2\r) \l(|h\ft\al_*|+|\hat{\goth{a}}|\r) \l(|h\ft\al_*|-|\hat{\goth{a}}|\r).
\eeq
The smallness comes from the last term. Indeed, the decay assumption \eqref{eq:alexpdecay} gives
$$
\begin{aligned}
&\sup_{p\in\R^d}||h\ft\al_*(p)|-|\hat{\goth{a}}(p)||
\leq \sup_{p\in\R^d}|h\ft\al_*(p)-\hat{\goth{a}}(p)|
 \leq \|h\al_*-\goth{a}\|_{L^1(\R^d)}\\
\leq &h\int_{B_{\phi(h)}^c} |\al_*(r)| \d r
=h\int_{B_{\phi(h)}^c} |\al_*(r)| e^{\rho_*r}e^{-\rho_*r}\d r \leq Ch e^{-\rho_* \phi(h)/2}.
\end{aligned}
$$
Note also that \eqref{eq:alexpdecay} implies $\|\ft\al_*\|_{L^\it(\R^d)}\leq\|\al_*\|_{L^1(\R^d)}\leq C$ and consequently $\|\ft{\goth{a}}\|_{L^\it(\R^d)}\leq Ch$. Applying these estimates to \eqref{eq:from}, we get
$$
|h\ft\al_*|^4-|\hat{\goth{a}}|^4\leq Ch^2 e^{-\rho_* \phi(h)/2} \l(|h\ft\al_*|^2+|\hat{\goth{a}}|^2\r).
$$
Recalling the definition \eqref{eq:gbcsgothadefn}, this implies
$$
\begin{aligned}
|g_{BCS}(\goth{a})-g_{BCS}(h\al_*)|
\leq &Ch^2 e^{-\rho_* \phi(h)/2} \int_{\R^d} (p^2+E_b) \l(|h\ft\al_*|^2+|\hat{\goth{a}}|^2\r) \d p\\
\leq &Ch^2 e^{-\rho_* \phi(h)/2} \l(h^2\|\al_*\|_{H^1(\R^d)}^2+\|\goth{a}\|_{H^1(\R^d)}^2\r).
\end{aligned}
$$
To conclude the claim \eqref{eq:chi2}, it remains to see that $\|\goth{a}\|_{H^1(\R^d)}^2\leq Ch^2$ as $h\downarrow 0$. For the $L^2$ part of the $H^1$ norm this follows from $\chi^2\leq 1$. For the derivative term, we denote $\chi_h\equiv \chi(\cdot/\phi(h))$ and use the Leibniz rule to get
$$
\|\nabla\goth{a}\|_{L^2(\R^d)}^2\leq 2h^2 \l(\|\chi_h\nabla\al_*\|_{L^2(\R^d)}^2+\|\al_*\nabla \chi_h\|_{L^2(\R^d)}^2\r).
$$
For the first term, we use $\chi^2\leq 1$ to get $\|\chi_h\nabla\al_*\|_{L^2(\R^d)}^2\leq \|\chi_h\nabla\al_*\|_{L^2(\R^d)}^2\leq C$. The second term is in fact much smaller:
\beq
\label{eq:fclaim}
\|\al_*\nabla\chi_h\|_{L^2(\R^d)}^2\leq Ce^{-2\rho_* \phi(h)}.
\eeq
Indeed, by H\"older's inequality and \eqref{eq:alexpdecay} we have
$$\begin{aligned}
\|\al_*\nabla\chi_h\|_{L^2(\R^d)}^2
=&\|\al_*\nabla\chi_h\|_{L^2(B_{2\phi(h)}\setminus B_{\phi(h)})}^2
\leq e^{-2\rho_*\phi(h)} \|\nabla\chi_h\|_{L^{\it}(\R^d)}^2\\
=& e^{-2\rho_*\phi(h)} \phi(h)^{-2} \|\nabla\chi\|_{L^\it(\R^d)}^2
\leq Ce^{-2\rho_* \phi(h)}.
\end{aligned}
$$
In the last step we used $\phi(h)\to\it$ as $h\to 0$. This proves \eqref{eq:fclaim} and completes the proof of \eqref{eq:chi2}. The argument for \eqref{eq:chi4} is even simpler.

Finally, we come to \eqref{eq:chi3}.  Since $(-\De+E_b+V)\al_*=0$, 
$$
\qmscp{\goth{a}}{-\De+E_b+V}{\goth{a}}=h\qmscp{\goth{a}}{[-\De,\chi_h]}{\al_*}=h^2 \|\al_*\nabla\chi_h\|_{L^2(\R^d)}^2.
$$
Therefore, \eqref{eq:chi3} follows from \eqref{eq:fclaim} and Proposition \ref{prop:chi} is proved.
\e{proof} 
%
%

\subsection{Conclusion}
Given $\psi\in H_0^1(\Om_{\ell(h)}^-)$, we extend it by zero to a function in $H^1(\R^d)$. Then, we define $\Gam_\psi$ as in Proposition \ref{prop:Gampsi}. We have
$$\begin{aligned}
\curly{E}^{BCS}(\Gam_\psi)=
&\Tr{\goth{h}\goth{a}_\psi\ol{\goth{a}_\psi}}+\iint_{\R^d\times \R^d}V\l(\frac{x-y}{h}\r) |\goth{a}_\psi(x,y)|^2 \d x\d y\\
&+ (1+h^{1/2}) \Tr{\goth{h}\goth{a}_\psi\ol{\goth{a}_\psi}\goth{a}_\psi\ol{\goth{a}_\psi}}.
\end{aligned}
$$
We apply the semiclassical expansion in Lemma \ref{lm:semiclassics} (note that the assumptions are satisfied by $\goth{a}$, since it is as regular as $\al_*$ and of compact support). We find, using $D=h^{-2}(\mu+E_b)$,
$$\begin{aligned}
&\curly{E}^{BCS}(\Gam_\psi)\\
= &h^{-d}\|\psi\|^2_{L^2(\R^d)} \qmscp{\goth{a}}{-\De+E_b+V}{\goth{a}}+ \|\goth{a}\|_{L^2(\R^d)}^2\l(\frac{h^{2-d}}{4}\|\nabla\psi\|^2_{L^2(\R^d)}-h^{2-d}D \|\psi\|^2_{L^2(\R^d)}\r)\\
&+ h^{2-d} \|\goth{a}\|_{L^2(\R^d)}^2\int_{\R^d} W(X) |\psi(X)|^2 \d X+h^{-d} g_{BCS}(\goth{a}) \|\psi\|_{L^4(\R^d)}^4\\
&+O(h^{5-d})(\|\psi\|_{H^1(\R^d)}^2+\|\psi\|_{H^1(\R^d)}^4)
\end{aligned}
$$
The main term in this expression is $h^{4-d}$ times the GP energy defined in \eqref{eq:EGPUdefn}, up to errors which are controlled by Proposition \ref{prop:chi} and the choice $\phi(h)=\log(h^{-q})$ with $q$ sufficiently large compared to $1/\rho_*$. We find
$$\begin{aligned}
\curly{E}^{BCS}(\Gam_\psi)=\curly{E}^{GP}_{\R^d}(\psi)+(O(h^{5-d})-Ch^{6-d}D)(\|\psi\|_{H^1(\R^d)}^2+\|\psi\|_{H^1(\R^d)}^4).
\end{aligned}
$$
Of course, $\curly{E}^{GP}_{\R^d}(\psi)=\curly{E}^{GP}_{\Om_{\ell(h)}^-}(\psi)$, since in fact $\psi\in H_0^1(\Om_{\ell(h)}^-)$.
This proves Theorem \ref{thm:key1} (UB).
\qed

\section{Proof of Theorem \ref{thm:key1} (LB): Decomposition}
\label{sect:LBA}
We prove Theorem \ref{thm:key1} (LB) and (LBC) together. (The situation will drastically simplify for convex $\Om$ in due course.) 

In this first part of the proof, we consider any BCS state $\Gam$ satisfying $\curly{E}^{BCS}(\Gam)\leq C_\Gam h^{4-d}$ (we think of $\Gam$ as an approximate BCS minimizer) and we show that its off-diagonal element $\al$ can be decomposed as in \eqref{eq:aldecomp}, with good a priori $H^1$ control on all the functions involved.


\be{thm}[Decomposition and a priori bounds]
Define 
$$
\tilde \Om:=\frac{\Om+\Om}{2}.
$$  
\label{thm:apriori}
Suppose that $\mu\leq -E_b+O(h^2)$ and that $\Gam$ is an admissible BCS state satisfying $\curly{E}^{BCS}(\Gam)\leq C_\Gam h^{4-d}$. Then, there exist $\psi\in H_0^1(\tilde\Om)$ and $\xi\in H_0^1(\tilde\Om\times\R^d)$ such that $\al$, the upper right entry of $\Gam$, can be decomposed as in \eqref{eq:aldecomp}. Moreover, we have the bounds
\beq
\label{eq:aprioribounds}
\begin{aligned}
&\|\nabla\psi\|_{L^2(\tilde\Om)}\leq C\|\psi\|_{L^2(\tilde\Om)}\leq O(1),\\
&\|\xi\|_{L^2(\tilde\Om\times\R^d)}^2+h^2 \|\nabla\xi\|_{L^2(\tilde\Om\times\R^d)}^2\leq O(h^{4-d}) (\|\psi\|_{L^2(\tilde\Om)}^2+C_\Gam).
\end{aligned}
\eeq
\e{thm}

The key input to the proof is the spectral gap of the operator $-\De+V$ above its ground state energy $-E_b$.
%

\subsection{Center of mass coordinates}
 
\be{lm}
\label{lm:apriori1}
Suppose that $\mu\leq -E_b+O(h^2)$. Let $\Gam$ be an admissible BCS state. Define the fiber
$$
\curly{D}:=\setof{(X,r)\in \tilde \Om\times \R^d}{X+\frac{r}{2}, X-\frac{r}{2}\in \Om}.
$$
Set $\tilde\al(X,r):=\al(X+r/2,X-r/2)$ so that $\tilde\al\in H_0^1(\curly{D})$. Then, for sufficiently small $h>0$, we have
$$
\begin{aligned}
\curly{E}^{BCS}(\Gam)\geq \iint\limits_{\curly{D}} \ol{\tilde \al(X,r)}\Bigg(&-\frac{h^2}{4}\Delta_X-h^2\Delta_r +h^2 W(X+r/2)-\mu\\
&+V(r/h)\Bigg) \tilde\al(X,r)\d r\d X+\frac{E_b}{2}\Tr{\al\ol{\al}\al\ol{\al}}.
\end{aligned}
$$
\e{lm}

We separate the following statement from the proof for later use
\be{prop}
\label{prop:gothh}
For $h$ small enough, $\goth{h}\geq E_b/2>0$.
\e{prop}

\be{proof}
By Assumption \ref{ass:V} $W$ is infinitesimally form-bounded with respect to $-\De_\Om$. Hence, $|W|\leq -\frac{1}{2} \Delta + C$ and $\goth{h}\geq -\frac{h^2}{2} \Delta-\mu-h^2 C$ hold in the sense of quadratic forms on $H_0^1(\Om)$. Since $\mu\leq -E_b+O(h^2)$, this implies that $\goth{h}\geq \frac{E_b}{2}$ for small enough $h$.
\e{proof}

We come to the

\be{proof}[Proof of Lemma \ref{lm:apriori1}]
The key input is that for any BCS state, we have the relations $\al\ol{\al}\leq \gam-\gam^2\leq \gam$ and we use the to pass from $\gam$ to $\al$ in the term $\tr{\goth{h}\gam}$. 
\beq
\label{eq:firsttwo}
\curly{E}^{BCS}(\Gam)
\geq 
\Tr{\goth{h}\al\ol{\al}}+\iint\limits_{\Om^2} V\l(\frac{x-y}{h}\r) |\al(x,y)|^2\d x\d y+\Tr{\goth{h}\gam^2}.
\eeq
We estimate the last term further. By Proposition \ref{prop:gothh}, $\al\ol{\al}\leq \gam$ and the fact that $A\mapsto \Tr{A^2}$ is operator monotone, we have
$$
\Tr{\goth{h}\gam^2}\geq \frac{E_b}{2} \Tr{\gam^2}\geq \frac{E_b}{2} \Tr{\al\ol{\al}\al\ol{\al}}.
$$
We now rewrite the first two terms in \eqref{eq:firsttwo} in center of mass coordinates. Using $\al(x,y)=\al(y,x)$ ($\Gam$ is Hermitian), we can write out the first term as
$$
\begin{aligned}
&\Tr{\goth{h}\al\ol{\al}}=\iint\limits_{\Om^2} \ol{\al(x,y)}\l(-h^2\Delta_x+h^2 W(x) -\mu+V\l(\frac{x-y}{h}\r)\r) \al(x,y)\d x\d y\\
=&\iint\limits_{\Om^2} \ol{\al(x,y)}\l(-\frac{h^2}{2}\Delta_x-\frac{h^2}{2}\Delta_y+h^2 W(x) -\mu+V\l(\frac{x-y}{h}\r)\r) \al(x,y)\d x\d y.
\end{aligned}
$$
Now we change to center-of-mass coordinates 
\beq
\label{eq:com}
X=\frac{x+y}{2},\qquad r=x-y, \qquad \tilde\al(X,r):=\al(X+r/2,X-r/2).
\eeq
Since the Jacobian is equal to one and $\Delta_x+\Delta_y=\frac{1}{2}\Delta_X+2\Delta_r$, Lemma \ref{lm:apriori1} follows.
\e{proof}

\subsection{Definition of the order parameter $\psi$}

An important idea is that from now on we isometrically embed $H_0^1(\curly{D})\subset H_0^1(\tilde \Om\times \R^d)$ by extending functions by zero. Note that all local norms are left invariant by the extension, in particular $\|\tilde\al\|_{L^2(\curly{D})}=\|\tilde\al\|_{L^2(\tilde \Om\times \R^d)}$.

We define the order parameter $\psi$ and establish some of its basic properties.

\be{prop}
\label{prop:psidefn}
Let $\tilde\al\in H_0^1(\curly{D})\subset H_0^1(\tilde \Om\times \R^d)$. For a fixed $X\in\tilde\Om$, we define the fiber
$$
\D_X:=\setof{r\in \R^d}{(X,r)\in \curly{D}}=\setof{r\in \R^d}{X+\frac{r}{2},X-\frac{r}{2}\in\Om}
$$
Let
\begin{align}
\label{eq:psidefn}
\psi(X):=&h^{-1}\int_{\D_X} \al_*(r/h)\tilde\al(X,r)\d r, \qquad &\textnormal{ for all } X\in\tilde\Om,\\
\label{eq:alpsidefn}
\tilde\al_\psi(X,r):=&h^{1-d}\psi(X)\al_*(r/h),\qquad &\textnormal{ for a.e. } X\in\tilde\Om,\, r\in\R^d,\\
\label{eq:xidefn}
\xi(X,r):=&\,\tilde\al(X,r)-\tilde\al_\psi(X,r),\qquad &\textnormal{ for a.e. } X\in\tilde\Om,\, r\in\R^d.
\end{align}
Then:
\be{enumerate}[label=(\roman*)]
\item $\psi\in H_0^1(\tilde\Om)$ and $\xi\in H_0^1(\tilde \Om\times \R^d)$. 
\item We have the norm identities
\beq
\begin{aligned}
\label{eq:orthosum}
\|\tilde\al\|_{L^2(\curly{D})}^2&=h^{2-d} \|\psi\|_{L^2(\tilde\Om)}^2+\|\xi\|_{L^2(\tilde \Om\times \R^d)}^2,\\
\|\nabla_X\tilde\al\|_{L^2(\curly{D})}^2&=h^{2-d} \|\nabla\psi\|_{L^2(\tilde\Om)}^2+\|\nabla_X\xi\|_{L^2(\tilde \Om\times \R^d)}^2.
\end{aligned}
\eeq
\e{enumerate}
\e{prop}

\be{proof}
From the definition of the weak derivative, we get that $\psi\in H_0^1(\tilde\Om)$ with 
\beq
\label{eq:nablapsi}
\nabla\psi(X)=h^{-1}\int_{\D_X} \al_*(r/h)\nabla_X\tilde\al(X,r)\d r.
\eeq
Since $\al_*\in H^1(\R^d)$ and $H_0^1(\tilde \Om\times \R^d)$ is a vector space, we also get $\xi\in H_0^1(\tilde \Om\times \R^d)$. This proves claim (i). For claim (ii), we observe the orthogonality relation
\beq
\label{eq:orthogonal}
\int_{\R^d} \al_*(r/h)\xi(X,r)\d r=0,
\eeq
which holds for a.e.\ $X\in\tilde\Om$. Thus, by expanding the square that one gets from \eqref{eq:xidefn} and using $\|\al_*(\cdot/h)\|^2_{L^2(\R^d)}=h^d$,
$$
\begin{aligned}
\|\tilde\al\|_{L^2(\curly{D})}^2=\|\tilde\al\|_{L^2(\tilde \Om\times \R^d)}^2=h^{2-d} \|\psi\|_{L^2(\tilde\Om)}^2+\|\xi\|_{L^2(\tilde \Om\times \R^d)}^2.
\end{aligned}
$$
This is the first identity in \eqref{eq:orthosum}. The second one follows by an analogous argument using \eqref{eq:nablapsi}.
\e{proof}

\subsection{Bound on the $W$ term}

\be{lm}
\label{lm:W}
Let $\tilde\al\in H_0^1(\curly{D})\subset H_0^1(\tilde \Om\times \R^d)$. For every $\eps>0$, there exists $C_\eps>0$ such that 
$$
\begin{aligned}
&\int_{\tilde\Om}\int_{\R^d}  |W(X+r/2)||\tilde\al_\psi(X,r)|^2\d r\d X
\leq h^{4-d}\l( \eps  \|\nabla\psi\|_{L^2(\tilde\Om)}^2+C_\eps\|\psi\|_{L^2(\tilde\Om)}^2\r)\\
&\int_{\tilde\Om}\int_{\R^d}  |W(X+r/2)||\xi(X,r)|^2\d r\d X
\leq  h^{2}\l( \eps \|\nabla\xi\|^2_{L^{2}(\tilde\Om \times\R^d)}+C_\eps\|\xi\|^2_{L^{2}(\tilde\Om \times\R^d)}\r).
 \end{aligned} 
$$
holds for sufficiently small $h$.
\e{lm}


\be{proof}
Recall that $\tilde\al=\tilde\al_{\psi}+\xi$, see \eqref{eq:xidefn}. In the following, we freely identify functions with their extensions by zero to all of $\R^d$, respectively to all of $\R^d\times\R^d$. By the semiclassical expansion in Lemma \ref{lm:semiclassics}(ii),
$$
\begin{aligned}
&\int_{\tilde\Om}\int_{\R^d}  |W(X+r/2)||\tilde\al_\psi(X,r)|^2\d r\d X\\
\leq&h^{2-d}\int_{\R^d} |W(X)| |\psi(X)|^2 \d X+C h^{3-d} \|W\|_{L^{p_W}(\R^d)} \|\psi\|_{H^1(\R^d)}^2\\
=&h^{2-d}\int_{\Om} |W(X)| |\psi(X)|^2\d X +C h^{3-d} \|W\|_{L^{p_W}(\Om)} \|\psi\|_{H_0^1(\tilde\Om)}^2.
\end{aligned}
$$
In the second step, we used our knowledge of where the functions are actually supported. Recall that $W$ is infinitesimally form-bounded with respect to $-\Delta$. Hence, for every $\eps>0$, there exists $C_\eps>0$ such that
$$
\int_{\Om} |W(X)| |\psi(X)|^2\d X\leq \eps \|\nabla\psi\|_{L^2(\Om)}^2+C_\eps \|\psi\|_{L^2(\Om)}^2
$$
This proves the first claimed bound.

By H\"older's inequality (on the space $\tilde\Om \times\R^d$ with Lebesgue measure) and the Sobolev interpolation inequality (on $\R^d\times\R^d$), we get that for every $\eps>0$, there exists $C_\eps>0$ such that
$$
\begin{aligned}
&\int_{\tilde\Om}\int_{\R^d}  |W(X+r/2)||\xi(X,r)|^2\d r\d X\\
\leq &2^{d/2} |\tilde\Om|^{1/2}\|W\|_{L^{2}(\Om)}  \|\xi\|^2_{L^{4}(\tilde\Om \times\R^d)}=2^{d/2} |\tilde\Om|^{1/2}\|W\|_{L^{2}(\Om)}  \|\xi\|^2_{L^{4}(\R^d\times\R^d)}\\
\leq &2^{d/2} |\tilde\Om|^{1/2}\|W\|_{L^{2}(\Om)}  \l(\eps \|\nabla\xi\|^2_{L^{2}(\tilde\Om \times\R^d)}+C_\eps \|\xi\|^2_{L^{2}(\tilde\Om \times\R^d)}\r)\\
\end{aligned} 
$$

Since $p_W\geq 2$ in all dimensions, this finishes the proof of Lemma \ref{lm:W}.
\e{proof}

\subsection{Proof of Theorem \ref{thm:apriori}}
The auxiliary results proved so far combine to give the following $H^1$ type lower bound on $\curly{E}^{BCS}$. From it, the a priori bounds stated in Theorem \ref{thm:apriori} will readily follow.

\be{lm}
\label{lm:H1}
Assume that $\mu\leq -E_b+O(h^2)$. Let $\tilde\al\in H_0^1(\curly{D})\subset H_0^1(\tilde\Om\times\R^d)$ be decomposed as $\tilde\al=\tilde\al_{\psi}+\xi$ as in Proposition \ref{prop:psidefn}. Then, there exist constants $c_1,c_2>0$ such that
$$
\begin{aligned}
\curly{E}^{BCS}(\Gam)\geq& c_1 h^2 \l(h^{2-d} \|\nabla\psi\|_{L^2(\tilde\Om)}^2+\|\nabla\xi\|_{L^2(\tilde \Om\times \R^d)}^2\r)+c_1\|\xi\|_{L^2(\tilde \Om\times \R^d)}^2\\
&-(\mu+E_b+c_2 h^2)\|\tilde\al\|_{L^2(\tilde\Om\times \R^d)}^2+\frac{E_b}{2}\Tr{\al\ol{\al}\al\ol{\al}}.
\end{aligned}
$$
holds for all sufficiently small $h$.
\e{lm}

\be{proof}
Given the bounds from Lemma \ref{lm:W} on the $W$ term, one can follow the proof of Lemma 3 in \cite{B}. The key ingredient is the spectral gap of the operator $-\De+V$ above its ground state (and the standard fact that the gap can be used to obtain $H^1$ control on the error term).
\e{proof}

\emph{Proof of Theorem \ref{thm:apriori}.}
Let $\mu\leq -E_b+O(h^2)$ and let $\Gam$ be a BCS state satisfying $\curly{E}^{BCS}(\Gam)\leq C_\Gam h^{4-d}$. By Lemma \ref{lm:H1} and $\mu\leq -E_b+O(h^2)$, we have
\beq
\label{eq:return}
\begin{aligned}
O(h^2)\|\tilde\al\|_{L^2(\tilde\Om\times \R^d)}^2
+C_\Gam h^{4-d}
\geq &h^2 \l(h^{2-d} \|\nabla\psi\|_{L^2(\tilde\Om)}^2+\|\nabla\xi\|_{L^2(\tilde \Om\times \R^d)}^2\r)\\
&+\|\xi\|_{L^2(\tilde \Om\times \R^d)}^2+\Tr{\al\ol{\al}\al\ol{\al}}
\end{aligned}
\eeq
We will eventually use all the terms in this equation. But first we note that \eqref{eq:return} gives 
\beq
\label{eq:xiH1}
\|\xi\|_{L^2(\tilde \Om\times \R^d)}^2\leq O(h^2)\|\tilde\al\|_{L^2(\tilde\Om\times \R^d)}^2+C_\Gam h^{4-d}.
\eeq From the first identity in \eqref{eq:orthosum}, we get 
$$\|\al\|_{L^2(\Om^2)}^2 \leq h^{2-d}\|\psi\|^2_{L^2(\tilde\Om)}+O(h^2)\|\al\|_{L^2(\Om^2)}^2+C_\Gam h^{4-d}$$ and so, for all sufficiently small $h$,
\beq
\label{eq:alpsiUB}
\|\al\|_{L^2(\Om^2)}^2\leq C h^{2-d}\|\psi\|^2_{L^2(\tilde\Om)}+C_\Gam h^{4-d}.
\eeq
Applying \eqref{eq:alpsiUB} to \eqref{eq:return} and dropping some non-negative terms, we conclude
\begin{align}
\label{eq:psinablapsi}
&\|\nabla\psi\|_{L^2(\tilde\Om)}^2\leq C( \|\psi\|_{L^2(\tilde\Om)}^2+C_\Gam),\\
&\|\xi\|_{L^2(\tilde\Om\times\R^d)}^2+h^2\|\nabla\xi\|_{L^2(\tilde\Om\times\R^d)}^2\leq O(h^{4-d})\l(\|\psi\|_{L^2(\tilde\Om)}^2+C_\Gam\r).
\end{align}
Thus, to prove \eqref{eq:aprioribounds}, it remains to show

\be{lm}
\label{lm:remains}
$\|\psi\|_{L^2(\tilde\Om)}\leq O(1)$. 
\e{lm}

\be{rmk}
\label{rmk:trgamma}
At this stage, \cite{B} prove Lemma \ref{lm:remains} (in three dimensions) by using $\|\psi\|^2_{L^2}\leq h\|\al\|_{L^2}^2=h\Tr{\al\ol{\al}}\leq h\Tr{\gam}$ and the fact that they work at fixed particle number $\Tr{\gam}=N/h$. Since we do not have this assumption, we use the semiclassical expansion of the quartic term $\Tr{\al\ol{\al}\al\ol{\al}}$. Here, as in the proof of Lemma \ref{lm:tildeOmLB} and in \cite{B}, one uses that in the Schatten norm estimate $\|\xi\|_{\goth{S}^4}\leq \|\xi\|_{\goth{S}^2}$, the right hand side is still of higher order in $h$ for dimensions $d\leq 3$.
\e{rmk}

\be{proof}[Proof of Lemma \ref{lm:remains}]
We retain only the trace on the right-hand side of \eqref{eq:return},
\beq
\label{eq:S4bound}
Ch^2\|\al\|_{L^2(\Om^2)}^2+C_\Gam h^{4-d}=Ch^2\|\tilde\al\|_{L^2(\tilde\Om\times\R^d)}^2 \geq \Tr{\al\ol{\al}\al\ol{\al}}.
\eeq
For the following argument, we extend all the relevant kernels to functions on $\R^d\times\R^d$. In this way, we can identify $\Tr{\al\ol{\al}\al\ol{\al}}\equiv \|\al\|^4_{\goth{S}^4}$, where $\|\cdot\|_{\goth{S}^p}$ denotes the Schatten trace norm of an operator on $L^2(\R^d)$. Equation \eqref{eq:xidefn} may be rewritten as
\beq
\begin{aligned}
\label{eq:splitting}
\al=\al_\psi+\tilde\xi,\qquad &\al_\psi(x,y)=h^{1-d}\psi\l(\frac{x+y}{2}\r)\al_*\l(\frac{x-y}{h}\r),\\
 &\tilde\xi(x,y)=\xi\l(\frac{x+y}{2},x-y\r).
\end{aligned}
\eeq
Here and in the following, the kernel functions $\al_\psi,\tilde\xi$ are understood to be functions on $\R^d\times \R^d$ (obtained by extension by zero). The Schatten norms satisfy the triangle inequality and are monotone decreasing in $p$. Also, the $\|\cdot\|_{\goth{S}^2}$ norm of any operator agrees with the $\|\cdot\|_{L^2(\R^d\times \R^d)}$ norm of its kernel. From these facts, we obtain
$$
\begin{aligned}
\|\al\|_{\goth{S}^4}
&\geq \|\al_\psi\|_{\goth{S}^4}-\|\tilde\xi\|_{\goth{S}^4}\geq \|\al_\psi\|_{\goth{S}^4}-\|\tilde\xi\|_{\goth{S}^2}=\|\al_\psi\|_{\goth{S}^4}-\|\tilde\xi\|_{L^2(\R^d\times\R^d)}\\
&=\|\al_\psi\|_{\goth{S}^4}-\|\xi\|_{L^2(\tilde\Om\times\R^d)}\geq \|\al_\psi\|_{\goth{S}^4}+O(h) \|\al\|_{L^2(\Om^2)}+O(h^{2-d/2}).
 \end{aligned}
$$
In the last step, we used \eqref{eq:xiH1}. From this, \eqref{eq:S4bound} and \eqref{eq:alpsiUB}, we get
\beq
\label{eq:above0}
\begin{aligned}
\|\al_\psi\|_{\goth{S}^4}^4
&\leq C\l(\|\al\|_{\goth{S}^4}^4+h^4 \|\al\|^4_{L^2(\Om^2)}+O(h^{8-2d})\r)\\
&\leq C \l(h^2\|\al\|^2_{L^2(\Om^2)}+h^4\|\al\|^4_{L^2(\Om^2)}+O(h^{4-d})\r)\\
&\leq C \l(h^{4-d}\|\psi\|^2_{L^2(\tilde\Om)}+h^{8-2d}\|\psi\|^4_{L^2(\tilde\Om)}+O(h^{4-d})\r).
\end{aligned}
\eeq
Along the way, we used $8-2d> 4-d$ for $d=1,2,3$. After extension by zero, $\psi\in H^1(\R^d)$ and we apply Lemma \ref{lm:semiclassics} (iv) to get
$$
\|\al_\psi\|_{\goth{S}^4}^4= h^{4-d} g_0(\al_*) \|\psi\|_{L^4(\tilde\Om)}^4+O(h^{5-d}) \|\psi\|_{H_0^1(\tilde\Om)}^4.
$$
Then, by \eqref{eq:psinablapsi} and H\"older's inequality, $\|\al_\psi\|_{\goth{S}^4}^4\geq C h^{4-d} \|\psi\|_{L^2(\tilde\Om)}^4$. Combining this estimate with \eqref{eq:above0} and using $8-2d> 4-d$, we get
$$
 \|\psi\|_{L^2(\tilde\Om)}^4
\leq C\|\psi\|^2_{L^2(\tilde\Om)}+O(1)
$$
This proves $\|\psi\|_{L^2(\tilde\Om)}\leq O(1)$ and hence Lemma \ref{lm:remains} and Theorem \ref{thm:apriori}.
\e{proof}

\section{Proof of Theorem \ref{thm:key1} (LB): Semiclassics}
\label{sect:LBB}

\subsection{From a priori bounds to GP theory}
We begin by deriving a lower bound in terms of GP energy on $\tilde \Om$, by assuming a decomposition with a priori bounds as in Theorem \ref{thm:apriori} and applying the semiclassical expansion from Lemma \ref{lm:semiclassics}.


Accordingly, in this section, $\psi$ and $\xi$ are general functions, not necessarily the ones defined previously in Proposition \ref{prop:existence} (they will be the same for convex domains).

\be{lm}
\label{lm:tildeOmLB}
Let $\mu\leq -E_b+O(h^2)$ and define $\nu':=\min\{d/2, 1\}$. Let $\Gam$ be a BCS state such that $\al$ can be decomposed as in \eqref{eq:aldecomp} for some $\psi\in H_0^1(\tilde \Om)$ and $\xi\in H_0^1(\tilde\Om\times\R^d)$. Moreover, suppose that $\|\psi\|_{H^1_0(\tilde \Om)}\leq O(1)$ and $\xi$ satisfies the bound in \eqref{eq:aprioribounds}.
Then, wee have
\beq
\label{eq:OmtildeLB}
\curly{E}^{BCS}(\Gam)
\geq h^{4-d} \curly{E}^{GP}_{\tilde \Om}(\psi) + O(h^{4-d+\nu'})\|\psi\|_{H^1_0(\tilde \Om)}^2,
\eeq
for some $C>0$. Here $\curly{E}^{GP}_{\tilde \Om}$ is defined with the parameter $D:=h^{-2}(\mu+E_b)$. 
\e{lm}

\subsubsection{Proof of Lemma \ref{lm:tildeOmLB}}
It will be convenient to define the auxiliary energy functional
$$
\begin{aligned}
\curly{E}_{LB}(\al):=&\Tr{(-h^2\Delta_\Om+h^2W-\mu)\al\ol{\al}}\\
&+\iint\limits_{\Om\times \Om} V\l(\frac{x-y}{h}\r) |\al(x,y)|^2\d x \d y
+ \Tr{\goth{h}\al\ol{\al}\al\ol{\al}}.
\end{aligned}
$$
We first note that this auxiliary functional provides a lower bound to the BCS energy. The basic idea is to replace $\gam$ by expressions in $\al$ using $\al\ol{\al}\leq \gam$ as in the proof of Lemma \ref{lm:apriori1}. However some additional difficulty is present here because the last term in $ \curly{E}_{LB}(\al)$ still features $\goth{h}$ and so we need the stronger operator inequality \eqref{eq:new} below.

\be{prop}
\label{prop:aux}
For sufficiently small $h$, we have $\curly{E}^{BCS}(\Gam)\geq \curly{E}_{LB}(\al)$, where $\al$ denotes the off-diagonal element of the BCS state $\Gam$.
\e{prop}

\be{proof}[Proof of Proposition]
The claim will follow from the operator inequality
\beq
\label{eq:new}
\gam\geq \al\ol{\al}+\al\ol{\al}\al\ol{\al}.
\eeq
To prove \eqref{eq:new}, we start by observing that $1-\ol{\gam}\leq (1+\ol{\gam})^{-1}$ by the spectral theorem. Consequently
$$
0\leq \Gam=\ttmatrix{\gam}{\al}{\ol{\al}}{1-\ol{\gam}}\leq \ttmatrix{\gam}{\al}{\ol{\al}}{(1+\ol{\gam})^{-1}}.
$$
The Schur complement formula implies
$$
\gam\geq \al(1+\ol{\gam})\ol{\al}.
$$ 
Using $\ol{\gam}\geq \ol{\al}\al$, we find
$$
\gam\geq \al(1+\ol{\gam})\ol{\al}\geq \al\ol{\al}+\al\ol{\al}\al\ol{\al}
$$
which proves \eqref{eq:new}. To conclude, let $h$ be sufficiently small such that $\goth{h}\geq 0$, see Proposition \ref{prop:gothh}. Then \eqref{eq:new} yields
$$
\Tr{\goth{h}\gam}\geq \Tr{\goth{h}\al\ol{\al}}+\Tr{\goth{h}\al\ol{\al}\al\ol{\al}}
$$ 
and this proves Proposition \ref{prop:aux}.
\e{proof}

The following key lemma says that we can apply the semiclassical expansion to the auxiliary energy functional with the desired result.

\be{lm}
\label{lm:auxiliary}
Under the assumptions of Lemma \ref{lm:tildeOmLB}, we use the splitting $\al=\al_\psi+\tilde\xi$ from \eqref{eq:splitting}. Then
$$
\curly{E}_{LB}(\al)\geq \curly{E}_{LB}(\al_\psi)+O(h^{4-d+\nu'})\|\psi\|_{H^1_0(\tilde\Omega)}^2.
$$
\e{lm}

Before we prove this lemma, we note that it directly implies Lemma \ref{lm:tildeOmLB}. Indeed, it gives
$$
\curly{E}^{BCS}(\Gam)\geq \curly{E}_{LB}(\al)\geq \curly{E}_{LB}(\al_\psi)+O(h^{4-d+\nu'})\|\psi\|_{H^1_0(\tilde\Omega)}^2.
$$
We extend $\psi\in H_0^1(\tilde\Omega)$ by zero to get an element of $H^1(\R^d)$, which we also denote by $\psi$. Then, all the terms in $\curly{E}_{LB}(\al_\psi)$ were computed in the semiclassical expansion in Lemma \ref{lm:semiclassics}. On the result of the expansion, we use the eigenvalue equation $(-\Delta+V+E_b)\al_*=0$ and recall $g_{BCS}(\al_*)=g_{BCS}$ from \eqref{eq:gbcsdefn}. This yields $\curly{E}^{GP}_{\R^d}(\psi)$ plus the error terms. These are as claimed, because $\|\psi\|_{H^1(\R^d)}\leq O(1)$ and $\mu\leq -E_b+O(h^2)$ by assumption. Finally, $\psi\in H_0^1(\tilde\Omega)$ implies $\curly{E}^{GP}_{\R^d}(\psi)=\curly{E}^{GP}_{\tilde\Omega}(\psi)$.

It remains to give the

\be{proof}[Proof of Lemma \ref{lm:auxiliary}]


We treat the terms in $\curly{E}_{LB}$ in four separate parts. First, by changing to center-of-mass coordinates \eqref{eq:com}, compare the proof of Lemma 3 in \cite{B},
\beq
\label{eq:remains}
\begin{aligned}
&\Tr{(-h^2\Delta_{\Om}+E_b)\al\ol{\al}}+\iint\limits_{\R^d\times\R^d} V\l(\frac{x-y}{h}\r) |\al(x,y)|^2\d x \d y\\
\geq & \Tr{(-h^2\Delta_{\Om}+E_b)\al_\psi\ol{\al_\psi}}+\iint\limits_{\R^d\times\R^d} V\l(\frac{x-y}{h}\r) |\al_\psi(x,y)|^2\d x \d y.
\end{aligned}
\eeq
Second, from $\mu\leq-E_b+O(h^2)$, \eqref{eq:alpsiUB} and \eqref{eq:aprioribounds}, we get
\beq
\label{eq:mu}
- (\mu+E_b)\Tr{\al\ol{\al}}\geq -(\mu+E_b) \Tr{\al_\psi\ol{\al_\psi}}+O(h^{6-d})\|\psi\|_{L^2(\tilde\Omega)}^2.
\eeq
Next, by Cauchy-Schwarz, Lemma \ref{lm:W} and \eqref{eq:aprioribounds}:
$$\begin{aligned}
\Tr{W\al\ol{\al}}
\geq &\Tr{W\al_\psi\ol{\al_\psi}}
-C \l(\|\xi\|_{L^2(\tilde\Omega\times\R^d)}^2+h^2\|\nabla\xi\|_{L^2(\tilde\Omega\times\R^d)}^2\r)\\
&-C\l(\|\xi\|_{L^2(\tilde\Omega\times\R^d)}^2+h^2\|\nabla\xi\|_{L^2(\tilde\Omega\times\R^d)}^2\r)^{1/2}
h^{1-\frac{d}{2}}\|\psi\|_{H_0^1(\tilde\Omega)}\\
\geq &\Tr{W\al_\psi\ol{\al_\psi}}+O(h^{3-d}).
\end{aligned}
$$
Using $\goth{h}=-h^2\Delta_{\Om}+h^2W-\mu$, the claim will then follow from
\beq
\label{eq:al4}
\Tr{\goth{h}\al\ol{\al}\al\ol{\al}}\geq \Tr{\goth{h}\al_\psi\ol{\al_\psi}\al_\psi\ol{\al_\psi}}+O(h^{4-d+\nu'}).
\eeq
This can be obtained by expanding the quartic and using the a priori bounds \eqref{eq:aprioribounds}, see the proof of (7.12) in \cite{B}. Here, we only explain how to treat the $W$ terms. Consider e.g.\ $\Tr{W\al_\psi\ol{\al}\al\ol{\tilde\xi}}$. By cyclicity of the trace, H\"older's inequality for Schatten norms and form-boundedness,
\beq
\label{eq:Wlater}
\begin{aligned}
\Tr{W\al_\psi\ol{\al}\al\ol{\tilde\xi}} 
\leq  &\|\al\|_{\goth{S}^6}^2\|\sqrt{|W|}\al_\psi\|_{\goth{S}^6} \|\sqrt{|W|}\mathrm{sgn}(W)\tilde\xi\|_{\goth{S}^2}\\
= &\|\al\|_{\goth{S}^6}^2\|\ol{\al_\psi}|W|\al_\psi\|_{\goth{S}^3}^{1/2}  \|\ol{\tilde\xi}|W|\tilde\xi\|_{\goth{S}^1}^{1/2}\\
 \leq & 2\|\al\|_{\goth{S}^6}^2\l(\|\nabla\al_\psi\|_{\goth{S}^6} \|\nabla\tilde\xi\|_{\goth{S}^2}+C\|\al_\psi\|_{\goth{S}^6} \|\tilde\xi\|_{\goth{S}^2}\r)   
\end{aligned}
\eeq
In the last step, we used the fact that form-boundedness can be stated as an operator inequality. When we multiply through by $h^2$, the last quantity is of the same form as the first term in (7.16) of \cite{B}. Using the same arguments as there with the a priori bounds \eqref{eq:aprioribounds} proves that it is $O(h^{4-d+\nu'})$. (We mention (a) that $\|\al\|_{\goth{S}^6}$ is estimated via $\|\al\|_{\goth{S}^6}\leq \|\al_\psi\|_{\goth{S}^6}+\|\tilde\xi\|_{\goth{S}^2}$, which is implicit in \cite{B} and (b) that the proof of Lemma 1 in \cite{B} generalizes to $d=1,2$ and gives $\|\al_\psi\|_{\goth{S}^6}\leq O(h^{1-d/6})$.)

 The same idea applies to all the other $W$ dependent terms in the expansion of the quartic. This proves Lemma \ref{lm:auxiliary} and consequently Lemma \ref{lm:tildeOmLB}.
\e{proof}

\subsection{Proof of Theorem \ref{thm:key1} (LBC)}
Let $\Om$ be convex and let $\Gam$ be an approximate BCS minimizer, i.e.\ $\curly{E}^{BCS}(\Gam)\leq C_\Gam h^{4-d}$. We apply Theorem \ref{thm:apriori} and then Lemma \ref{lm:tildeOmLB}. Since $\Om=\tilde\Om$ by convexity, this finishes the proof.
\qed

\subsection{Proof of Theorem \ref{thm:key1} (LB)}
\label{sect:LBnonC}
Let $\Om$ be a non-convex bounded Lipschitz domain. The order parameter $\psi$ defined in Proposition \ref{prop:psidefn} now lives on $\tilde \Om=\frac{\Om+\Om}{2}$, which may be a much larger set than $\Om$. 

\subsubsection{Decay of the order parameter}
We first show that $\psi$ in fact \emph{decays exponentially} away from $\Om$. This follows easily from its definition \eqref{eq:psidefn} and the exponential decay of $\al_*$, see \eqref{eq:alexpdecay}.

 \be{prop}
\label{prop:psipwise}
There exists a constant $C_0>0$ such that for every $\ell>0$ and almost every $X\in\tilde\Om$ with $\dist(X,\Om)\geq \ell$, we have
\begin{align}
\label{eq:psipwiseestimate}
|\psi(X)|&\leq C_0 h^{d/2-1}e^{-\rho_*\frac{2\ell}{h}} \|\tilde\al(X,\cdot)\|_{L^2(\curly{D}_X)}\\
\label{eq:nablapsipwiseestimate}
|\nabla\psi(X)|&\leq C_0 h^{d/2-1}e^{-\rho_*\frac{2\ell}{h}} \|\nabla_X\tilde\al(X,\cdot)\|_{L^2(\curly{D}_X)}.
\end{align}
\e{prop}

\be{proof}
Let $\ell>0$ and $X\in\tilde\Om$ with $\dist(X,\Om)\geq \ell$. The key observation is that the triangle inequality implies
$$
\curly{D}_X\subseteq \setof{r\in\R^d}{|r|> 2\ell},
$$
where $\curly{D}_X$ was defined in Proposition \ref{prop:psidefn}. Therefore, by Cauchy-Schwarz and \eqref{eq:alexpdecay}
$$
\begin{aligned}
|\psi(X)|
&\leq h^{-1}\int_{\D_X} |\al_*(r/h)||\tilde\al(X,r)|\d r\\
&= h^{-1}  \int_{\D_X} e^{-\rho_*\frac{r}{h}}e^{\rho_*\frac{r}{h}} |\al_*(r/h)| |\tilde\al(X,r)|\d r\\
&\leq C_0 h^{d/2-1} e^{-\rho_*\frac{2\ell}{h}} \|\tilde\al(X,\cdot)\|_{L^2(\curly{D}_X)}. 
\end{aligned} 
$$
This proves \eqref{eq:psipwiseestimate}. Starting from \eqref{eq:nablapsi}, the same argument gives \eqref{eq:nablapsipwiseestimate}.
\e{proof}

\subsubsection{Conclusion by a cutoff argument}
With Proposition \ref{prop:psipwise} at our hand, we just have to cut off part of $\psi$ that lives sufficiently far away from $\Om$.
We first apply Theorem \ref{thm:apriori} to get the decomposition and the a priori bounds stated there. Then, we define
$$
\begin{aligned}
\psi_1(X):&=\eta_{\frac{\ell(h)}{4},\Om_{\ell(h)}^+}(X)\psi(X),\\
 \xi_1(X,r):&=\xi(X,r)+(\psi(X)-\psi_1(X))\al_*(r/h).
\end{aligned}
$$
Here $\Om^+_{\ell}$ was defined in \eqref{eq:Omell+defn}, the cutoff function $\eta_{\ell,U}$ was defined in \eqref{eq:etadefn} and $\ell(h)=h\log(h^{-q})$. Note that we also have \eqref{eq:aldecomp} with $\psi,\xi$ replaced by $\psi_1,\xi_1$. 

Note that $\psi_1\in H_0^1(\Om_{\ell(h)}^+)$ and consequently 
$$
\curly{E}^{GP}_{\tilde\Om}(\psi_1)=\curly{E}^{GP}_{\Om_{\ell(h)}^+}(\psi_1).
$$
Thanks to this, the claim will follow from Lemma \ref{lm:tildeOmLB} applied with the choices $\psi=\psi_1$, $\xi=\xi_1$. It remains to show that its assumptions are satisfied, namely that $\|\psi_1\|_{H^1_0(\Om_{\ell(h)}^+)}\leq O(1)$ and $\xi_1$ satisfies \eqref{eq:aprioribounds}.\\ 

For this part, we denote $\eta\equiv \eta_{\frac{c_0\ell(h)}{4},\Om^+_{\ell(h)}}$ and $\ell\equiv \ell(h)$ for short. We first prove that $\|\psi_1\|_{H^1_0(\Om^+_\ell)}\leq O(1)$. Using $\eta\leq 1$ and Cauchy-Schwarz, we get
\beq
\label{eq:psi1}
\|\psi_1\|_{H^1_0(\Om^+_\ell)}^2\leq 2 \|\psi\|_{H^1_0(\Om^+_\ell)}^2 + 2 \int_{\Om^+_\ell(h)} |\nabla \eta|^2 |\psi|^2 \d X
=O(1) + 2 \int_{\Om^+_\ell} |\nabla \eta|^2 |\psi|^2 \d X.
\eeq
The term with $|\nabla\eta|$ may look troubling since we can only control $|\nabla\eta|\leq \ell^{-2}$ on $\supp\,\nabla \eta$. The key insight is that this potential blow up in $h$ is sufficiently dampened on $\supp\,\nabla \eta$ by the exponential decay of $|\psi|$ established by Proposition \ref{prop:psipwise}. Namely, we will prove 

\be{lm}
\label{lm:eta}
$\supp\,\nabla\eta(p)\subset (\Om_{\ell/2}^+)^c$
\e{lm}
We postpone the proof of this geometrical lemma for now. Assuming it holds, it is straightforward to use the decay estimates from Proposition \ref{prop:psipwise} to conclude from \eqref{eq:psi1} that $\|\psi_1\|_{H^1_0(\Om^+_\ell)}\leq O(1)$, by choosing $q$ large enough (with respect to $1/\rho_*$). 

Next, we show that $\xi_1$ satisfies $\eqref{eq:aprioribounds}$. From Theorem \ref{thm:apriori}, we already know that $\xi$ satisfies \eqref{eq:aprioribounds}. When integrating the other term in the definition of $\xi_1$, we change to center of mass coordinates and write $\psi-\psi_1=\psi(1-\eta)$. Since $\nabla(1-\eta)$ and $\nabla\eta$ are supported on the same set, one can use the argument from above again on the center of mass integration (i.e.\ a combination of Lemma \ref{lm:eta} and Proposition \ref{prop:psipwise}). We leave the details to the reader. 

To finish the proof of Theorem \ref{thm:key1} (LB), it remains to give the

\be{proof}[Proof of Lemma \ref{lm:eta}]
Let $p\in \R^d$ be a point such that $\nabla\eta(p)\neq 0$. Then, by definition \eqref{eq:etadefn} of $\eta$,
$$
\dist(p,(\Om_\ell^+)^c)< \ell/2.
$$
Let $q_\ell\in (\Om_\ell^+)^c$ be a point such that $\dist(p,(\Om_\ell^+)^c)=|p-q_\ell|$ and let $q\in \ol{\Om}$ be a point such that $\dist(p,\Om)=|p-q|$ (such points exists by a compactness argument). By definition \eqref{eq:Omell+defn} of $\Om_\ell^+$ and the triangle inequality,
$$
\ell\leq \dist(\Om,(\Om_{\ell}^+)^c)\leq |q-q_\ell|\leq |q-p|+|p-q_\ell|< |q-p|+\ell/2.
$$
Therefore, $\dist(p,\Om)=|q-p|> \ell/2$ and so $p\in (\Om_{\ell/2}^+)^c$. Since $p$ was an arbitrary point with $\nabla\eta(p)\neq 0$ and $(\Om_{\ell/2}^+)^c$ is closed, Lemma \ref{lm:eta} is proved.
\e{proof}

\section{Proof of the continuity of the GP energy (Theorem \ref{thm:key2})}
\label{sect:key2}

\subsection{Davies' use of Hardy inequalities}
\label{sect:hardy}
This section serves as a preparation to prove the second key result Theorem \ref{thm:key2}.

The central idea that we discuss here is Lemma \ref{lm:Davies}. It is based on the insight of Davies \cite{Davies93, Davies00} that continuity of the Dirichlet energy under interior approximations of a domain $U$ follows from good control on the boundary decay of functions that lie in the operator domain of $\De_U$, under the sole assumption that the domain $U$ satisfies a Hardy inequality \eqref{eq:hardy}.

Importantly, \emph{GP minimizers for $\curly{E}^{GP}_U$ are in $\mathrm{dom}(\De_U)$ thanks to the Euler Lagrange equation}, see Proposition \ref{prop:existence}, and this is how one derives the continuity of the GP energy (Theorem \ref{thm:key2}) from Lemma \ref{lm:Davies}. 

As its input, the lemma requires the validity of the

\be{defn}[Hardy inequality]
Let $U\subseteq \R^d$ and denote
\beq
\label{eq:dudefn}
d_{U}(x):=\mathrm{dist}(x,U^c).
\eeq We say that $U$ satisfies a Hardy inequality, if there exist $c_U\in(0,1]$ and $\lam\in \R$ such that
\beq
\label{eq:hardy}
\int_U d_{U}(x)^{-2} |\vp(x)|^2 \d x\leq \frac{4}{c_U^2} \|\nabla \vp\|^2_{L^2(U)}+\lam \|\vp\|^2_{L^2(U)},\qquad \forall \vp\in C_c^\it(U).
\eeq
We shall refer to $c_U$ and $\lam$ as the ``Hardy constants''.
\e{defn}

We can now state

\be{lm}
\label{lm:Davies}
For any $0<\ell<1$, we define the function $\eta_{\ell,U}:\R^d\to [0,\it)$ by
\beq
\label{eq:etadefn}
\eta_{\ell,U}(x):=\be{cases}
0,\qquad &\text{if }0\leq d_{U}(x)\leq \ell\\
\frac{d_{U}(x)-\ell}{\ell},\qquad &\text{if }\ell\leq d_{U}(x)\leq 2\ell\\
1,\qquad &\text{otherwise}.
\e{cases}
\eeq 
Suppose that $U$ satisfies the Hardy inequality \eqref{eq:hardy} for some $c_U\in (0,1]$ and some $\lam\in \R$. Then, there exists a constant $c>0$ depending only on $c_U$ and $\lam$ such that 
$$
\curly{E}^{GP}_U(\eta_{\ell,U} \vp)-\curly{E}^{GP}_U(\vp)\leq c \ell^{c_U} \l(\|\vp\|_{H_0^1(U)}\|\De_U \vp\|_{L^2(U)}+\|\vp\|_{H_0^1(U)}^2\r),\quad \forall \vp\in\mathrm{dom}(\De_U).
$$
Moreover, the same bound holds for the quantity $\|\eta_{\ell,U} \vp\|^2_{H^1_0(U)}-\|\vp\|^2_{H^1_0(U)}$.
\e{lm}

We remark that
$\eta_{\ell,U}$ is a Lipschitz continuous function with a Lipschitz constant that is independent of $U$ (this is because $d_U$ has the Lipschitz constant one for all $U$).

\be{proof}
We write $\eta\equiv \eta_{\ell,U}$. First, we note that the nonlinear term drops out because $|\eta \vp|^4-|\vp|^4=(\eta^4-1)|\vp|^4\leq 0$ thanks to $0\leq \eta\leq 1$. 
For the gradient term, we note that the Hardy inequality \eqref{eq:hardy} is the main assumption in \cite{Davies93, Davies00}. Thus, by Lemma 11 in \cite{Davies00}, there exists a $c>0$ (depending only on the Hardy constants $c_U$ and $\lam$) such that
$$
\int_U (|\nabla(\eta \vp)|^2-|\nabla \vp|^2)\d x\leq c\ell^{c_U} \|\De_U\vp\|_{L^2(U)} \|\nabla \vp\|_{L^2(U)}, \quad \forall \vp\in\mathrm{dom}(-\De_U).
$$
Since $\eta\leq 1$, this already implies the last sentence in Lemma \ref{lm:Davies}.
Using Cauchy-Schwarz, Assumption \ref{ass:V} on $W$ and  Theorem 4 in \cite{Davies00}, we get
$$
\begin{aligned}
&\int_U (W+D) (\eta^2-1)|\vp|^2 \d x
\leq \int_U (|W|+|D|) (1-\eta^2)|\vp|^2 \d x\\
 \leq
&\l(\|W\vp\|_{L^{2}(\Om)}+|D|\|\vp\|_{L^{2}(\Om)}\r) \l(\int_{U\cap \{d_{U}\leq 2\ell\}} |\vp|^2 \d x\r)^{1/2}\\
\leq &c \l(\|W\|_{L^{p_W}(\Om)}+|D|\r) \|\vp\|_{H_0^1(U)} \ell^{1+c_U/2} \l(\|\De_U\vp\|_{L^2(U)} \|\nabla \vp\|_{L^2(U)}\r)^{1/2}
\end{aligned}
$$
for another constant $c$ depending only on $c_U$ and $\lam$. We estimate the last term via $2\sqrt{ab}\leq a+b$. Then we use that $\ell^{1+c_U/2}\leq \ell^{c_U}$ holds for all $c_U\in (0,1]$ and $0<\ell<1$. This proves \ref{lm:Davies}.
\e{proof}

With Lemma \ref{lm:Davies} at our disposal, we need conditions on $U$ such that it satisfies the Hardy inequality \eqref{eq:hardy}. 

In a fundamental paper, Necas \cite{Necas} proved that any bounded Lipschitz domain $\Om$ satisfies a Hardy inequality for some $c_\Om\in (0,1]$ and some $\lam\in \R$. Hence, we can apply Lemma \ref{lm:Davies} with $U=\Om$ and this is already sufficient to obtain continuity of the GP energy under \emph{interior} approximation, i.e.\ Theorem \ref{thm:key2} with $\Om_{\ell}^-$. Hence, Necas' result is already sufficient to derive

\be{itemize}
\item[(i)] the upper bounds in the two main results, Theorems \ref{thm:main1} and \ref{thm:main2}.
\item[(ii)] the complete Theorem \ref{thm:main2} for bounded and \emph{convex} domains $\Om$. Indeed, Theorem \ref{thm:key1} (LBC) gives the lower bound and the upper bound holds because any convex domains satisfies a Hardy inequality \cite{Marcusetal, MS}. (In fact, the Hardy constants can be taken as $c=4$ and $\lam=0$.)
\e{itemize}

The continuity of the GP energy under \emph{exterior} approximation (and therefore our proof of the lower bounds in the main results) relies on the following new theorem. It is is an extension of Necas' argument \cite{Necas}. The proof is deferred to Appendix \ref{app:hardy}.

\be{thm}
\label{thm:hardy}
Let $\Om$ be a bounded Lipschitz domain. There exist $c_\Om\in (0,1]$, $\lam\in \R$ and $\ell_0>0$, as well as a sequence of exterior approximations $\{\Om_\ell\}_{0<\ell<\ell_0}$ such that the Hardy inequality \eqref{eq:hardy} holds with $U=\Om_{\ell}$ for all $\ell<\ell_0$. 

Moreover, the sequence of approximations $\{\Om_\ell\}_\ell$ satisfies the following properties.
\be{enumerate}[label=(\roman*)]
\item There exists a constant $c_0>1$ such that $\Om_\ell^+\subset \Om_\ell\subset \Om_{c_0\ell}^+$.
\item There exists a constant $a>0$ such that
\beq
\label{eq:containment}
\setof{q\in\R^d}{\dist(q,(\Om_\ell)^c)>a\ell}\subset \Om.
\eeq
\e{enumerate}
\e{thm}

We emphasize that the Lipschitz character of $\Om$ is important for the sequence of approximations $\{\Om_\ell\}_\ell$ to exist. Concretely, properties (i) and (ii) cannot both hold for exterior approximations of the slit domain example presented in Remark \ref{rmk:example} (while there do exist approximations that all satisfy the Hardy inequality with the $\ell$-independent constant $c_\Om=1/2$).

\subsection{Proof of Theorem \ref{thm:key2}}
We begin by observing that $\Om_\ell^-\subset \Om\subset \Om_\ell^+$ trivially gives
$$
E^{GP}_{\Om_\ell^+}\leq E^{GP}_{\Om}\leq E^{GP}_{\Om_\ell^-}.
$$
Theorem \ref{thm:key2} says that the reverse bounds hold as well, up to the claimed error terms. The basic idea is to take a minimizer on the larger domain and to cut it off near the boundary, where the energy cost of the cutoff is controlled by Lemma \ref{lm:Davies}. 

\subsubsection{Interior approximation}
The situation is easier for interior approximation, since then we consider GP minimizers and the Hardy inequality on the fixed domain $\Om$. We want to apply Lemma \ref{lm:Davies} and we gather prerequisites.

First, by Proposition \ref{prop:existence}, there exists a unique non-negative minimizer for $\curly{E}^{GP}_{\Om}$, call it $\psi$, and it satisfies 
\beq
\label{eq:ELGbound}
\|\Delta_U\psi\|_{L^2(U)}\leq C(1+|D|) (\|\psi\|_{H^1_0(U)}+\|\psi\|_{H^1_0(U)}^3)
\eeq
Second, since $\Om$ is a bounded Lipschitz domain, there exist $c_\Om\in(0,1]$ and $\lam\in\R$ such that the Hardy inequality \eqref{eq:hardy} holds on $U=\Om$ \cite{Necas}.
Now we apply Lemma \ref{lm:Davies} with the domain $U=\Om$ and the cutoff function $\eta_{2\ell,\Om}$. We get
$$
\begin{aligned}
\curly{E}_{\Om}^{GP}(\eta_{2\ell,\Om}\psi)
\leq &\curly{E}_{\Om}^{GP}(\psi)+O(\ell^{2/c})(\|\psi\|_{H_0^1(\Om)}\|\De_\Om \psi\|_{L^2(\Om)}+\|\psi\|_{H_0^1(\Om)}^2)\\
\leq &\curly{E}_{\Om}^{GP}(\psi)+O(\ell^{2/c})
\end{aligned}
$$
In the second step, we used \eqref{eq:ELGbound} and the fact that all norms of $\psi$ are independent of $\ell$. The definitions of $\eta_{2\ell,\Om}$ and $\Om_\ell^-$ are such that $\supp\,\eta_{2\ell,\Om}\subset \Om_\ell^-$. Since $\eta_{2\ell,\Om}$ is Lipschitz continuous, this implies $\eta_{2\ell,\Om}\psi\in H_0^1(\Om_\ell^-)$ and therefore
\beq
\label{eq:asin}
\curly{E}_{\Om}^{GP}(\eta_{2\ell,\Om}\psi)=\curly{E}_{\Om_\ell^-}^{GP}(\eta_{2\ell,\Om}\psi)\geq E^{GP}_{\Om_\ell^-}.
\eeq
This proves the claimed continuity under interior approximation.

\subsubsection{Exterior approximation}
The idea is similar as before, but additional $\ell$ dependencies complicate the argument somewhat. We let $\{\Om_\ell\}_{0<\ell<\ell_0}$ be the sequence of exterior approximations given by Theorem \ref{thm:hardy}. That is, $\Om^+_\ell\subset\Om_{\ell}$ and the Hardy inequality \eqref{eq:hardy} holds on all $U=\Om_\ell$ with Hardy constants that are uniformly bounded in $\ell$.

By Proposition \ref{prop:existence}, there exists a unique non-negative minimizer for $\curly{E}^{GP}_{\Om_\ell}$, call it $\psi_\ell$, and it satisfies the analogue of \eqref{eq:ELGbound} with a $C$ that is independent of $\ell$.

Recall definition \eqref{eq:etadefn} of the cutoff function $\eta_{a\ell,\Om_\ell}$. Here we choose $a>0$ such that property (ii) in Theorem \ref{thm:hardy} holds which is equivalent to 
\beq
\label{eq:etasupport}
\supp\,\eta_{a\ell,\Om_\ell}\subset \Om.
\eeq
Now we apply Lemma \ref{lm:Davies}. We note that the constant $c$ appearing in it depends only on the Hardy constants (and these are uniformly bounded in $\ell$). Therefore, using the analogue of \eqref{eq:ELGbound}, we get
\beq
\label{eq:considering}
\curly{E}^{GP}_{\Om_\ell}(\eta_{a\ell,\Om_\ell}\psi_{\ell})\leq \curly{E}^{GP}_{\Om_{\ell}}(\psi_\ell) +O(\ell^{2/c}) O(\|\psi_{\ell}\|^2_{H^1_0(\Om_{\ell})}+\|\psi_{\ell}\|^4_{H^1_0(\Om_{\ell})}).
\eeq
Regarding the error term, we note

\be{lm}
\label{lm:psi+}
$\|\psi_{\ell}\|_{H^1_0(\Om_{\ell})}\leq O(1)$.
\e{lm}

\be{proof}[Proof of Lemma \ref{lm:psi+}]
We use that the GP energy can only increase under a decrease of the underlying domain to get
\beq
\label{eq:contradicts}
\curly{E}^{GP}_{\Om_{\ell}}(\psi_{\ell})=E^{GP}_{\Om_\ell}
\leq E^{GP}_{\Om}
\eeq 
The claim now follows from the coercivity  \eqref{eq:coercivity}, since the constants $C_1,C_2,D$ there do not depend on the underlying domain and hence not on $\ell$.  
\e{proof}

By \eqref{eq:etasupport} and the fact that $\eta_{a\ell,\Om_\ell}$ is a Lipschitz function, we get $\eta_{a\ell,\Om_\ell}\psi_{\ell}\in H_0^1(\Om)$. Returning to \eqref{eq:considering}, we can conclude the proof as in \eqref{eq:asin}, which yields Theorem \ref{thm:key2}.
\qed

\section*{Acknowledgements}
The authors would like to thank Christian Hainzl and Robert Seiringer for helpful discussions. R.L.F.\ was supported by the U.S. National Science Foundation through grants PHY-1347399 and DMS-1363432. B.S.\ was supported by the U.S.\ National Science Foundation through grant DMS-1265592 and by the Israeli Binational Science Foundation through grant 2014337.

\be{appendix}

\section{On GP minimizers}
\label{app:existence}
We prove Proposition \ref{prop:existence}.

\dashuline{Proof of (i).}
The coercivity \eqref{eq:coercivity} is a straightforward consequence form-boundedness of $W$ and the elementary bound 
$$
|\psi|^4-(C+D)|\psi|^2\geq -(C_2+D)^2.
$$
The constants $C_1,C_2$ only depend on $W$ and since $W$ is supported in the fixed domain $\Om$, they are independent of $U$. Clearly \eqref{eq:coercivity} implies $E_U^{GP}>-\it$. 

\dashuline{Proof of (ii).}
Let $\{\psi_n\}$ be a minimizing sequence for $E_U^{GP}$. By the coercivity \eqref{eq:coercivity}, the sequence is bounded in $H_0^1(U)$ and hence weakly $H_0^1(U)$-precompact. Let $\psi_*\in H_0^1(U)$ denote one of its weak limit points. After extending all functions by zero to get functions on $\R^d$, we obtain weak convergence in $H^1(\R^d)$. Therefore, by Rellich's theorem, $\psi_n\to \psi_*$ in all $L^p(U)$ spaces with $p< p_S(d)$ the critical Sobolev exponent in dimension $d$ (e.g.\ $p_S(3)=6$). Letting $p'$ denote the H\"older dual of $p$, we get
$$
\begin{aligned}
\int_U W                                                                                                                                                                                                                                                                                                                                                                                                                                                                                                                                                                                                                                                                                                          (|\psi_n|^2-|\psi_*|^2)\d x\leq &\l(\|W\psi_n\|_{L^{p'}(U)}+\|W\psi_*\|_{L^{p'}(U)}\r) \||\psi_n|-|\psi_*|\|_{L^p(U)}\\
\leq &C\|W\|_{L^{2}(U)} (\|\nabla\psi_n\|_{H_0^1(U)}+\|\nabla\psi_*\|_{H_0^1(U)}) \|\psi-\psi_*\|_{L^p(U)}\to 0.
\end{aligned}
$$
The last estimate holds by Assumption \ref{ass:V} for all $p<p_S(d)$ sufficiently close to $p_S(d)$. The same argument gives the continuity of the $D$ term in $\curly{E}^{GP}_U$.

Let ${\#}\in \{n,*\}$. We write $\curly{E}^{GP}_U(\psi_{\#})=A_{\#}+B_{\#}$, where $A_{\#}=\|\nabla\psi_{\#}\|_{L^2(U)}^2$ and $B_{\#}$ contains the remaining terms. Then, the above shows that $B_n\to B_*$. Moreover, by weak convergence is $H^1_0(U)$, $\liminf A_n \geq A_*$, so $E_U^{GP} = \lim (A_n+ B_n) \geq A_* + B_*$. Since $A_* + B_* \geq E_U^{GP}$ by definition of $E_U^{GP}$, we conclude that $\psi_*$ is a minimizer and that $A_n\to A_*$. Thus, $\|\psi_n\|_{H^1_0(U)}\to\|\psi_*\|_{H^1_0(U)}$ and therefore $\psi_n\to\psi_*$ strongly in $H_0^1(U)$.

To prove the uniqueness statement we first note that $\|\nabla |\psi|\|_{L^2(U)}\leq \|\nabla \psi\|_{L^2(U)}$. Moreover, since $\rho\mapsto \|\nabla \sqrt\rho\|_{L^2(U)}^2$ is convex and $\rho\mapsto \|\rho\|_{L^2(U)}^2$ is strictly convex, we see that $\mathcal E_D^{GP}(\psi)$ is a strictly convex functional of $|\psi|^2$, and therefore has a unique minimizer.

\dashuline{Proof for (iii).}
We compute the Euler Lagrange equation for the GP energy and find
$$
-\frac{1}{4}\Delta_U\psi_*+(W-D)\psi_* +2g_{BCS} |\psi_*|^2\psi_*=0.
$$
This equation holds in the dual of $H^1_0(U)$, that is, when tested against $H^1_0(U)$ functions. By our Assumption \ref{ass:V} on $W$ and Sobolev's inequality, $\Delta_U\psi_*$ is in fact an $L^2(U)$ function and we have the bound
$$
\begin{aligned}
\|\Delta_U\psi_*\|_{L^2(U)}=&\|4(W-D)\psi_* +8g_{BCS}|\psi_*|^2\psi_*\|_{L^2(U)}\\
 \leq& C(1+|D|) (\|\psi_*\|_{H^1_0(U)}+\|\psi_*\|_{H^1_0(U)}^3).
\end{aligned}
$$
This finishes the proof of Proposition \ref{prop:existence}.
\qed

\section{Convergence of the one body density}
\label{app:wc}
\be{proof}[Proof of Proposition \ref{prop:wc}]
We fix a real valued $w\in L^{p_W}(\Om)$ and $t\in \R$ and define $W_t:=W+tw$. We denote the BCS/GP energies which are defined with $W_t$ by $\curly{E}_t^{BCS},E^{BCS}_t,\curly{E}_t^{GP},$ etc. On the one hand, our assumption on $\Gam$ gives
$$
E^{BCS}-E_t^{BCS}\geq \curly{E}_t^{BCS}(\Gam)-\curly{E}_t^{BCS}(\Gam)+o(h^{4-d})=th^2 \Tr{\gam w}+o(h^{4-d}).
$$
On the other hand, Theorem \ref{thm:main2} yields
$$
E^{BCS}-E_t^{BCS}=h^{4-d} (E^{GP}-E_t^{GP})+O(h^{4-d+\nu})
$$
where the implicit constant depends on $w$. We denote the unique non-negative minimizer of $\curly{E}_t^{GP}$ by $\psi_t$ (see Proposition \ref{prop:existence}). Multiplying through by $h^{d-4}$ and taking $h\to0$, we find
\beq
\label{eq:t}
\limsup_{h\to 0} th^{d-2}\Tr{\gam w}\leq E^{GP}-E_t^{GP}\leq \curly{E}^{GP}(\psi_t)-\curly{E}_t^{GP}(\psi_t)=t \int_{\Om} w |\psi_t|^2\d x. 
\eeq
We claim that $\psi_t\to \psi_0\equiv \psi_*$ in $H_0^1(\Om)$. This will imply the main claim \eqref{eq:weakconv}. To see this, one divides \eqref{eq:t} by $t$, distinguishing the cases $t>0$ and $t<0$, and sends $t\to 0$. Then one uses Rellich's theorem to get $|\psi_t|^2\to |\psi_0|^2$ in $L^{p_W'}(\Om)$.

Hence, it remains to prove that $\psi_t\to \psi_*$ in $H_0^1(\Om)$. This is a simple compactness argument. We denote $\eta_t:=\psi_t-\psi_*$. The coercivity \eqref{eq:coercivity} and the triangle inequality imply that $\|\eta_t\|_{H_0^1(\Om)}$ remains bounded as $t\to 0$. We have
$$
\begin{aligned}
0\leq& \curly{E}^{GP}(\psi_t)-\curly{E}^{GP}(\psi_*)
=\curly{E}_t^{GP}(\psi_t)-\curly{E}_t^{GP}(\psi_*)-t\int_\Om w(2\Re (\eta_t)\psi_*+|\eta_t|^2)\d x\\
\leq& -t\int_\Om w(2\Re (\eta_t)\psi_*+|\eta_t|^2)\d x
\end{aligned}
$$
The right hand side vanishes as $t\to 0$, since $\|\eta_t\|_{H_0^1(\Om)}$ remains bounded as $t\to 0$. Therefore, $\psi_t$ is a sequence of approximate minimizers of $\curly{E}^{GP}$. Proposition \ref{prop:existence} (ii) then implies that $\psi_t\to \psi_*$ in $H_0^1(\Om)$.
\e{proof}

\section{On the semiclassical expansion}
\label{app:semiclassics}
We sketch the proof of Lemma \ref{lm:semiclassics}, especially where it departs from similar results in \cite{B}.
All norms and all integrals are taken over $\R^d$, unless noted otherwise.

\begin{proof}[Proof of Lemma \ref{lm:semiclassics}]
\emph{Proof of (i).} This follows directly from changing to the center-of-mass coordinates \eqref{eq:com}, compare the proof of Lemma \ref{lm:apriori1}. 

\emph{Proof of (ii).} We write out the trace with operator kernels, change to center-of-mass coordinates \eqref{eq:com} and apply the fundamental theorem of calculus to get
$$
\begin{aligned}
\Tr{W\goth{a}_\psi\ol{\goth{a}_\psi}} = &h^{-d} \iint W(X) |\goth{a}(r)|^2\l|\psi\l(X-\frac{hr}{2}\r)\r|^2 \d X\d r\\
 = & h^{-d}  \int W(X) |\psi(X)|^2 \d X-h^{-d} \eta
\end{aligned}
$$
with 
\beq
\label{eq:semi}
\eta= \mathrm{Re} \iint W(X)|\goth{a}(r)|^2\l(\int_0^1 \psi\l(X-\frac{shr}{2}\r) \scpp{hr}{\nabla\psi\l(X-\frac{shr}{2}\r)}	\d s\r) \d X\d r.
\eeq
By H\"older's and Sobolev's inequalities, $|\eta|\leq h\|W\|_{L^{p_W}(\Om)} \|\sqrt{|\cdot|} \goth{a}\|_{L^{2}}^2 \|\psi\|^2_{H^1}$. This is $O(h)$, since $\|\sqrt{|\cdot|} \goth{a}\|_{L^{2}}^2<\it$ by our assumptions on $\goth{a}$.

\emph{Proof of (iii).} 
The argument in Lemma 1 in \cite{B} generalizes because the critical Sobolev exponent is always greater or equal to six in $d=1,2,3$ and so all the error terms can be bounded in terms of $\| \psi\|_{H^1(\R^d)}$. 
We mention that the idea of the proof is to write the trace in terms of operator kernels and to change to the four body center-of-mass coordinates
$$
X=\frac{x_1+x_2+x_3+x_4}{4},\qquad r_k=x_{k+1}-x_k,\quad k=1,2,3.
$$
Then, one rescales the relative coordinates $r_k$ by $h$ (since they appear as $\goth{a}(r_k/h)$) and expands in $h$.

When proving the first equation in (iii), the $W$ term requires a different argument. Namely, as in the proof of \eqref{eq:al4}, one uses H\"older's inequality for Schatten norms and form-boundedness of $W$ with respect to $-\De$ to get
$$
|\Tr{W\al_\psi\ol{\al_\psi}\al_\psi\ol{\al_\psi}}| \leq C \l(\|\nabla\al_\psi\|_{\goth{S}^4}^2+\|\al_\psi\|_{\goth{S}^4}^2\r) \|\nabla\al_\psi\|_{\goth{S}^4}^2.
$$
Afterwards, one multiplies by $h^2$ and uses the bounds from Corollary 1 in \cite{B}. This gives the first equation in (iii). For the second equation in (iii), one replaces $\|V\goth{a}\|_{L^1}$ in the estimate of the error term $A_2$ in \cite{B} by $\|\goth{a}\|_{L^1}$, which is also finite. 
\e{proof}

\section{On Lipschitz domains and Hardy inequalities}
\label{app:hardy}
We first present the construction of a suitable sequence of exterior approximations to a bounded Lipschitz domain. Then, we prove that this sequence satisfies Hardy inequalities with uniformly bounded Hardy constants (Theorem \ref{thm:hardy}).

 The proof of Theorem \ref{thm:hardy} is an extension of Necas' argument \cite{Necas} for a fixed Lipschitz domain and draws on known results on the geometry of the sequence of the exterior approximations \cite{Calderon, Lipschitz}. (We remark that we could alternatively work with the naive enlargements $\Om_\ell^+$ \eqref{eq:Omelldefn}, but this would require writing down a non trivial amount of elementary geometry estimates.)


\subsection{Definitions}

We begin by recalling

\be{defn}[Lipschitz domain]
\label{defn:lipschitz}
A set $\Om\subseteq \R^d$ is a bounded Lipschitz domain, if its boundary $\del\Om$ can be covered by finitely many bounded and open coordinate cylinders $\curly{C}_1,\ldots,\curly{C}_K\subset \R^d$ such that for all $1\leq k\leq K$, there exist $R_k,\beta_k>0$ such that
$$
\begin{aligned}
\del\Om\cap \curly{C}_k=&\{(\x,f_k(\x))\in B_{R_k}\times \R\},\\
\Om\cap \curly{C}_k=&\setof{(\x,y)\in B_{R_k}\times \R}{-\beta_k<y<f_k(\x)},\\
\Om^c\cap \curly{C}_k=&\setof{(\x,y)\in B_{R_k}\times \R}{f_k(\x)<y<\beta_k}.
\end{aligned}
$$
where $f_k:B_{R_k}\to \R$ is a uniformly Lipschitz continuous function on $B_{R_k}\subset\R^{d-1}$, the ball of radius $R_k$ centered at the origin.

The equalities above should be understood in the sense that there exists an isometric bijection between the two sets (so the Cartesian coordinate frame defined through the $\x\in B_{R_k}$ may be different for each $k$).
\e{defn}

The exterior approximations $\Om_\ell$ are obtained by extending $\Om$ in the direction of a smooth transversal vector field, which any Lipschitz domain is known to host. 

\be{prop}[Normal and transversal vector fields]
\label{prop:transversal}
Let $\Om$ be a bounded Lipschitz domain in the sense of Definition \ref{defn:lipschitz}. For every $1\leq k\leq K$, define the outward normal vector field (to $\del\Om$) in the coordinate cylinder $\curly{C}_k$ by
\beq
n(\x):=\frac{(\nabla f_k(\x),-1)}{\sqrt{1+|\nabla f_k(\x)|^2}},
\eeq
for Lebesgue almost every $\x\in B_{R_k}$ and extend the definition to all $\x\in B_{R_k}$ by averaging over a small ball as in (2.4) of \cite{Lipschitz}.

Then, $\Om$ hosts a smooth vector field $v:\R^d\to \R^d$ which is ``transversal'', i.e.\ there exists $\kappa\in (0,1)$ such that for all $1\leq k\leq K$,
\beq
v(\x,f_k(\x))\cdot n(\x)\geq \kappa,\qquad |v(\x,f_k(\x))|=1,
\eeq
for almost every $\x\in B_{R_k}$.
\e{prop}

In the definition of $n(\x)$, we used the fact that the Lipschitz continuous function $f_k$ is differentiable almost everywhere by Rademacher's theorem.

The basic idea for Proposition \ref{prop:transversal} is that in each coordinate cylinder $\curly{C}_k$ from Definition \ref{defn:lipschitz}, one takes the constant vector field $e_d$, i.e.\ the $y$ direction, and then one smoothly interpolates between different $\curly{C}_k$ via a partition of unity. For the details, see e.g.\ pages 597-599 in \cite{Lipschitz} (and note that the surfaces measure, called $\sigma$ there, and the Lebesgue measure on $B_{R_k}$ are mutually absolutely continuous).

We are now ready to give

\be{defn}[Exterior approximations]
\label{defn:Omell}
Let $\Om$ be a bounded Lipschitz domain and let $v$ be the transversal vector field from Proposition \ref{prop:transversal}. For every $\ell>0$, define its enlargement by
\beq
\label{eq:Omelldefn}
\hat\Om_\ell:=\setof{p+\ell v(p)}{p\in \Om}.
\eeq
\e{defn}

\subsubsection{Bounds on $\hat\Om_\ell$}
Each set $\hat\Om_\ell$ has many nice properties if $\ell$ is small enough, see Proposition 4.19 in \cite{Lipschitz} (though this is stated for the case $\ell<0$, analogous results hold for $\ell>0$, as is also mentioned there). In particular, $\hat\Om_\ell$ is also a bounded Lipschitz domain and there exist coordinate cylinders in which both $\del\Om$ and $\del\hat\Om_\ell$ are represented as the graphs of Lipschitz continuous functions, with Lipschitz constants that are uniformly bounded in $\ell$. Moreover:

\be{prop}
\label{prop:Omellbounds}
There exists a constant $c_0>0$, such that for all $\ell>0$ small enough,
\beq
\label{eq:Omellbounds}
\Om_{c_0\ell}^+ \subset \hat\Om_\ell\subset \Om^+_\ell.
\eeq
\e{prop}

This lemma will give property (i) in Theorem \ref{thm:hardy}, up to reparametrizing it as $\Om_\ell:=\hat\Om_{\ell/c_0}$.

\be{proof}
The second containment follows directly from Proposition 4.15 in \cite{Lipschitz}.

For the first containment, we invoke Proposition 4.19 in \cite{Lipschitz}. It gives $\ol{\Om}\subset \hat\Om_\ell$ and consequently
\beq
\label{eq:distanceeq}
\dist(\Om,\hat\Om_\ell^c)=\dist(\del\Om,\hat\del\Om_\ell).
\eeq
We will show that $\dist(\del\Om,\del\hat\Om_\ell)\geq c_0\ell$. By Proposition 4.19 (i) in \cite{Lipschitz}, 
\beq
\label{eq:dellOm}
\del\hat\Om_\ell=\setof{p+\ell v(p)}{p\in \del\Om}.
\eeq
Hence, by a compactness argument, there exist $p,p'\in\del\Om$ such that
$$
\dist(\del\Om,\del\hat\Om_\ell)=|p'-(p+\ell v(p))|=|V(p',0)-V(p,\ell)|,
$$
where we introduced the map
\beq
\label{eq:Vdefn}
\begin{aligned}
V:\del\Om\times (-\ell_0,\ell_0)&\to\R^d\\
(p, s)&\mapsto p+s v(p).
\end{aligned}
\eeq
By (4.67) in \cite{Lipschitz}, $V$ is bi-Lipschitz if $\ell_0>0$ is small enough. In particular, there exists $c_0>0$ such that
$$
|V(p',0)-V(p,\ell)|\geq c_0|(p',0)-(p,\ell)|\geq c_0\ell.
$$
This proves $\dist(\del\Om,\del\hat\Om_\ell)\geq c_0\ell$. The claim then follows from \eqref{eq:distanceeq} and definition \eqref{eq:Omell+defn} of $\Om^+_\ell$.
\e{proof}

\subsubsection{Proof of Theorem \ref{thm:hardy}}

We apply Necas' proof \cite{Necas} to all $\Om_\ell$ simultaneously (with $\ell$ sufficiently small) and observe that all the relevant constants can be bounded uniformly in $\ell$.

By Proposition 4.19 (ii) in \cite{Lipschitz}, for $\ell_0>0$ small enough, there exist coordinate cylinders $\curly{C}_1,\ldots,\curly{C}_K$ that (a) cover $\del\Om_\ell$ for all $0\leq \ell<\ell_0$ and (b) characterize them as the graph of Lipschitz functions $f_{k,\ell}$ in the $e_d$ direction, as described in Definition \ref{defn:lipschitz}. Moreover, the Lipschitz constants of $f_{k,\ell}$ are uniformly bounded in $\ell$. 

Let $\curly{C}_0\subset\Om$ be an open set such that $\dist(\curly{C}_0,\Om^c)>0$ and such that $\Om\subset  \bigcup_{k=0}^K\curly{C}_k$. Let $\phi_0,\ldots,\phi_K:\R^d\to\R^d$ be a smooth partition of unity subordinate to this covering, i.e.
$$
\supp\, \phi_k\subset \curly{C}_k,\qquad \sum_{k=0}^K\phi_k=1 \text{ on } \bigcup_{k=0}^K\curly{C}_k.
$$
The key observation is that, locally, the distance $d_\ell:=\dist(\cdot,\del\Om_\ell)$ is comparable to $f_{k,\ell}-y$ up to constants which depend on the Lipschitz constant of $f_{k,\ell}$ and are thus uniformly bounded in $\ell$. Concretely, we have

\be{lm}
\label{lm:comparable}
There exist constants $a>0$ and $0<b\leq 1$ such that for all $1\leq k\leq K$ and all $0\leq \ell<\ell_0$, we have
\beq
\label{eq:comparable}
\min\{a,b|f_k(\x)-y|\}\leq d_\ell(\x,y)\leq |f_{k,\ell}(\x)-y|
\eeq
for all $(\x,y)\in \supp\phi_k$.
\e{lm}

\be{proof}
Fix $1\leq k\leq K$. The second inequality is trivial because $(\x,f_{k,\ell}(\x))\in \del\Om_\ell$ implies
$$
d_\ell(\x,y)\leq |(\x,y)-(\x,f_{k,\ell}(\x))|=|f_{k,\ell}(\x)-y|.
$$
For the proof of the first inequality in \eqref{eq:comparable}, we define
$$
a:=\min_{k=0,\ldots,K} \dist(\supp\,\phi_k,\del \curly{C}_k^c)>0.
$$
Since $\del\Om_\ell$ is compact, $d_\ell(\x,y)$ is achieved at some point $p_0\in\del\Om_\ell$. In case $p_0\not\in \curly{C}_k$, we can bound
$$
d_\ell(\x,y)=|p_0-(\x,y)|\geq a,
$$
and in case $p_0\in \curly{C}_k$ we can write it as $p_0=(\x_0,f_{k,\ell}(\x_0))$ and proceed as follows. Recall that every $f_{k,\ell}$ is Lipschitz continuous with a Lipschitz constant that is uniformly bounded in $\ell$; call the bound $L$. Hence, for every $\tau\in(0,1)$,
$$
\begin{aligned}
d_\ell(\x,y)^2=&(\x-\x_0)^2+(y-f_{k,\ell}(\x_0))^2\\
\geq &(\x-\x_0)^2+(1-\tau^{-1})(f_{k,\ell}(\x)-f_{k,\ell}(\x_0))^2+(1-\tau)(y-f_{k,\ell}(\x_0))^2\\
\geq &(1-L(\tau^{-1}-1))(\x-\x_0)^2+(1-\tau)(y-f_{k,\ell}(\x))^2.
\end{aligned}
$$
Now one chooses $\tau\in (0,1)$ so that $1-L(\tau^{-1}-1)=0$. This yields the first inequality in Lemma \ref{lm:comparable} with an appropriate $b>0$. We have thus proved Lemma \ref{lm:comparable}.
\e{proof}

We resume the proof of Theorem \ref{thm:hardy}. Take any $\vp\in C_c^\it(\Om_\ell)$ and use the partition of unity to write the left hand side of the Hardy inequality \eqref{eq:hardy} as
$$
\begin{aligned}
\int_{\Om_\ell} |\vp(x)|^2 d_\ell(x)^{-2}\d x
=&\sum_{k=0}^K \int_{\curly{C}_k\cap \Om_\ell} \phi_k(x)|\vp(x)|^2 d_\ell(x)^{-2}\d x\\
 \leq& C \|\vp\|_{L^2}^2 +\sum_{k=1}^K\int_{\curly{C}_k\cap \Om_{\ell_0}} \phi_k(x)|\vp(x)|^2 d_\ell(x)^{-2}\d x.
\end{aligned}
$$
where $C=\dist(\curly{C}_0,\Om^c)^{-2}<\it$. We emphasize that we used $\Om_{\ell}\subset\Om_{\ell_0}$ in the last integral. Now, we write each integral over $\curly{C}_k$ in boundary coordinates and apply Lemma \ref{lm:comparable}. Importantly, the resulting expression is independent of $\ell$ (it only depends on $\ell_0$). Hence, one can conclude the proof, exactly as in \cite{Necas}, by Fubini and the one-dimensional Hardy inequality \cite{HLP}. This proves the first part of Theorem \ref{thm:hardy}.\\

It remains to show properties (i) and (ii). (i) holds by Proposition \ref{prop:Omellbounds}. For (ii), we take any $q\in \R^d$ such that $\dist(q,\Om_\ell^c)\geq a\ell$. In particular, $q\in \Om_\ell$. Hence, if $\ell$ is small enough, there exists $p\in \Om$ such that
$$
q=p+\ell v(p).
$$
Recall that the vector field $v:\R^d\to\R^d$ is differentiable. We introduce the finite and $\ell$ independent constants
$$
C_0:=\|v\|_{L^\it(\ol{\Om_{\ell_0}})},\qquad C_1:=\|\nabla v\|_{L^\it(\ol{\Om_{\ell_0}})}.
$$
Using the characterization \eqref{eq:dellOm} and $q\in\Om_\ell$, we have
$$
\begin{aligned}
a\ell\leq& \dist(q,\Om_\ell^c)=\min_{p'\in\del\Om}|p+\ell v(p)-p'-\ell v(p')|\\
\leq& (1+C_1\ell) \min_{p'\in\del\Om} |p-p'|
=(1+C_1\ell)\dist(p,\Om^c).
\end{aligned}
$$
We can choose $\ell$ small enough so that $C_1\ell\leq 1$ (this uses that $C_1$ can only decrease if $\ell_0$ decreases).
We get
$$
\begin{aligned}
\dist(q,\Om^c)=&\inf_{p'\in\Om^c}|p+\ell v(p)-p'|
\geq  \inf_{p'\in\Om^c}|p-p'|-C_0\ell\\
=&\dist(p,\Om^c)-C_0\ell
\geq \ell (a/2-C_0).
\end{aligned}
$$
By choosing $a>0$ large enough, we get that $q\in \Om$ as claimed.
This finishes the proof of Theorem \ref{thm:hardy}.
\qed

%
%

\section{The linear case: Ground state energy of a two body operator}
\label{app:linear}

In this section, we discuss a linear version of our main result. It gives an asymptotic expansion of the ground state energy of the two body operator \eqref{eq:twobody}, describing a fermion pair which is confined to $\Om$ 

While in principle the center of mass and relative coordinate are coupled due to the boundary conditions, the result shows that they contribute to the ground state energy of $H_h$ on different scales in $h$ (and therefore in a decoupled manner).

\be{thm}
\label{thm:twobody}
Let $\Om\subset \R^d$ be a bounded Lipschitz domain. Given functions $V:\R^d\to \R$ and $W:\Om\to \R$ satisfying Assumption \ref{ass:V}, we define the two body operator
\beq
\label{eq:twobody}
H_h:=\frac{h^2}{2}(-\De_{\Om,x}+W(x)-\De_{\Om,y}+W(y))+V\l(\frac{x-y}{h}\r)
\eeq
with form domain $H_0^1(\Om\times \Om)$. Then, as $h\downarrow 0$,
\beq
\label{eq:linearclaim}
\inf\mathrm{spec}_{L^2(\Om\times \Om)} H_h=-E_b+h^2 D_c+O(h^{2+\nu}),
\eeq
where $\nu>0$ is as in Theorem \ref{thm:main2} (i) and
$$-E_b=\inf\mathrm{spec}_{L^2(\R^d)} (-\De+V),\qquad D_c=\inf\mathrm{spec}_{L^2(\Om)}\l(-\frac{1}{4}\De_{\Om}+W\r).$$
\e{thm}

This could be proved by following the line of argumentation in the main text and ignoring the nonlinear terms throughout. However, the proof of the lower bound is considerably simpler in the linear case, because a monotonicity argument eliminates the need for a priori bounds. To not obscure the key ideas, we give the proof in the special case when $W\equiv 0$ and $\Om$ is convex. 

It is instructive to think of the even more special case when $\Om$ is an interval, say $\Om=[0,1]$. This case is depicted in Figure \ref{fig:diamond} and the proof is sketched in the caption. 

 \begin{figure}[t]
\begin{center}
\includegraphics[height=9cm]{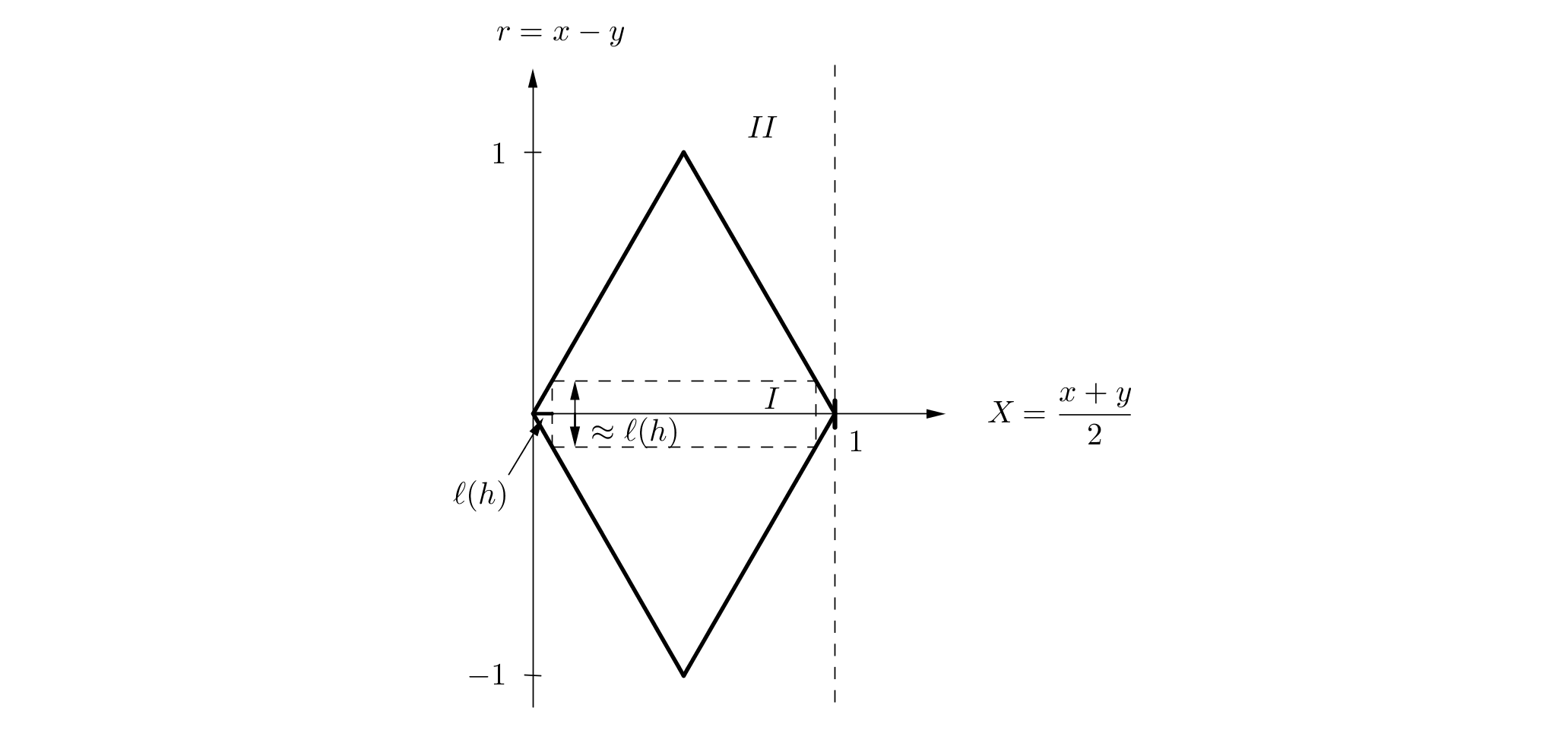}
\caption{When $\Om=[0,1]$, the region $\Om^2$ has a diamond shape when depicted in the center of mass coordinates $(X,r)$. To prove the upper bound in Theorem \ref{thm:twobody}, one uses a trial state, see \eqref{eq:tbtrial}, which is mostly (up to exponentially small errors in the $r$ direction) supported on the small dashed rectangular region $I$, where $\ell(h)=h\log(h^{-q})$ with $q>0$ large but fixed. When $\Om=[0,1]$, the Dirichlet eigenfunctions are explicit sine functions and so one does not need to invoke Theorem \ref{thm:key2} to get the upper bound. For the lower bound, one drops the Dirichlet condition in the relative variable, i.e.\ one extends the problem from the diamond to the strip $II=[0,1]\times \R$. This decouples the $X$ and $r$ variables and directly yields the lower bound.}
\label{fig:diamond}
\end{center}
\end{figure}

\be{proof}
We denote the ground state energy of $-\frac{1}{4}\De_{\Om_\ell^-}$ by $D_c^-(\ell)$ (compare \eqref{eq:Dcelldefn}), where $\Om_\ell^-$ is defined in \eqref{eq:Omell-defn}. 

\dashuline{Upper bound.} We construct a trial state with the following functions: $\al_*$, the ground state satisfying $(-\De+V)\al_*=-E_b\al_*$, $\chi$ a cutoff function as described in Definition \ref{defn:trial}, and $\psi_{\ell(h)}$, the normalized ground state of $-\De_{\Om_{\ell(h)}^-}$ for $\ell(h)=h\log(h^{-q})$ and $q>0$ large but fixed. In center of mass variables, $X=\frac{x+y}{2}$, $r=x-y$, the trial state then reads
\beq
\label{eq:tbtrial}
\psi_{\ell(h)}(X)\chi\l(\frac{r}{\ell(h)}\r)h^{1-d}\al_*\l(\frac{r}{h}\r).
\eeq
We apply $H_h$ to this and use the fact that $-\frac{1}{2}\De_x-\frac{1}{2}\De_y=-\frac{1}{4}\De_X-\De_r$. The exponential decay of $\al_*$ controls the localization error introduced by $\chi$ as in the proof of Proposition \ref{prop:chi}. Therefore the energy of the trial state is 
$-E_b+h^2 D^-_c(\ell(h))+O(h^{2+\nu})$. The second (linear) part of Theorem \ref{thm:key2} with $W\equiv 0$ says that $D^-_c(\ell(h))\leq D_c+O(h^\nu)$. Hence the upper bound in \eqref{eq:linearclaim} is proved.

\dashuline{Lower bound.} The key idea is to drop the Dirichlet boundary condition in the relative variable. The center of mass coordinates are originally defined on the domain
$$
\curly{D}:=\setof{(X,r)\in \Om\times \R^d}{X+ \frac{r}{2}, X-\frac r2\in \Om}.
$$
Observe that $\curly{D}\subset \Om\times \R^d$. On the space $L^2(\Om\times \R^d)$, we define a new operator 
$$
\tilde H_h=-\frac{h^2}{4}\De_{\Om,X}-h^2 \De_r+V(r/h),
$$
with form domain $H_0^1(\Om\times \R^d)$. By domain monotonicity we have $\tilde H_h\leq H_h$ in the sense of quadratic forms, and therefore
\beq
\label{eq:LBH}
\inf\mathrm{spec}_{L^2(\Om\times \R^d)} \tilde H_h\leq \inf\mathrm{spec}_{L^2(\Om\times \Om)} H_h.
\eeq
Now $\inf\mathrm{spec}_{L^2(\Om\times \R^d)} \tilde H_h$ can be computed exactly since the variables are decoupled. The ground state is just
$$
\psi_0(X)h^{1-d}\al_*\l(\frac{r}{h}\r)
$$
where $\psi_0$ is the normalized ground state of $-\frac{1}{4}\De_\Om$. The energy of this state is precisely equal to $-E_b+h^2 D_c$. By \eqref{eq:LBH}, the lower bound follows.
\e{proof}

\e{appendix}

\bibliographystyle{amsplain}

\begin{thebibliography}{10}


\bibitem{Ancona86}
A.~Ancona, \emph{{On strong barriers and an inequality of Hardy for domains in $R^n$}}, J. London Math. Soc. (2) \textbf{34} (1986), no. 2, 274-–290.

\bibitem{BLS94}
V.~Bach, E.~H. Lieb, and J.~P.~Solovej, \emph{{Generalized
  Hartree-Fock theory and the Hubbard model}}, J. Stat. Phys. \textbf{76}
  (1994), no.~1-2, 3--89 .

\bibitem{BCS57}
J.~Bardeen, L.~N. Cooper, and J.~R. Schrieffer, \emph{{Theory of
  Superconductivity}}, Phys. Rev. \textbf{108} (1957), 1175--1204.

\bibitem{B}
G.~Br\"aunlich, C.~Hainzl and R.~Seiringer \emph{{Bogolubov-Hartree-Fock theory for strongly interacting fermions in the low density limit}}, Math. Phys. Anal. Geom. 19 (2016), no. 2, 19:13.

\bibitem{BrezisMarcus}
H.~Brezis and M.~Marcus, \emph{{Hardy's inequalities revisited.}} Annali della Scuola Normale Superiore di Pisa - Classe di Scienze \textbf{25}.1-2 (1997): 217-237. 

\bibitem{Calderon}
A.P.~Calderon, \emph{Boundary value problems for the Laplace equation in Lipschitzian domains}, Recent progress in Fourier analysis (El Escorial, 1983), 33–48, North-Holland Math. Stud., 111, North-Holland, Amsterdam, 1985

\bibitem{Davies93}
E.~B.~Davies, \emph{{Eigenvalue stability bounds via weighted Sobolev spaces}} Math. Z. \textbf{214} (1993), no. 3, 357–371.

\bibitem{Davies00}
E.~B.~Davies, \emph{{Sharp boundary estimates for elliptic operators}}, Math. Proc. Cambridge Philos. Soc. \textbf{129} (2000), no. 1, 165-–178




\bibitem{deGennesBC}
P.G. de~Gennes, \emph{Boundary Effects in Superconductors}, Rev. Mod. Phys. 36, 225

\bibitem{deGennes}
P.G. de~Gennes, \emph{{Superconductivity of Metals and Alloys}}, Westview
  Press, 1966.
  
  \bibitem{DZ92}
  M. Drechsler and W. Zwerger, \emph{{Crossover from BCS-superconductivity to Bose-condensation}}, Annalen der Physik
\textbf{504} (1), 15-–23, (1992)

\bibitem{Evansetal}
W. D. Evans, D.J. Harris, and R. M. Kauffman, \emph{{Boundary behaviour of {D}irichlet eigenfunctions of second
              order elliptic operators}}, Math. Z., \textbf{204} (1990), no. 1, 85--15

\bibitem{FHNS07}
R.~L.~Frank, C.~Hainzl, S.~Naboko and R.~Seiringer,
  \emph{{The critical temperature for the BCS equation at weak coupling.}}
J. Geom. Anal. \textbf{17} (2007), no. 4, 559–567

\bibitem{FHSS12}
R.~L.~Frank, C.~Hainzl, R.~Seiringer, and J.~P.~Solovej,
  \emph{{Microscopic derivation of Ginzburg-Landau theory}}, J. Amer. Math.
  Soc. \textbf{25} (2012), 667--713.

\bibitem{FHSS12c}
R.~L.~Frank, C.~Hainzl, R.~Seiringer, and J.~P.~Solovej,
  \emph{{Derivation of Ginzburg-Landau theory for a one-dimensional system with contact interaction}}, Operator methods in mathematical physics, 57–88, Oper. Theory Adv. Appl., 227, Birkhäuser/Springer Basel AG, Basel, 2013
  
  \bibitem{FHSS15}
R.~L.~Frank, C.~Hainzl, R.~Seiringer, and J.~P.~Solovej,
  \emph{{The external field dependence of the BCS critical temperature}}, Comm. Math. Phys. 342 (2016), no. 1, 189 - 216
  
    \bibitem{FL15}
R.~L.~Frank, M.~Lemm,
  \emph{{Multi-Component Ginzburg-Landau Theory: Microscopic Derivation and Examples}}, Ann. Henri Poincar\'e,  DOI 10.1007/s00023-016-0473-x

\bibitem{Griffiths}
R.G.~Griffiths, \emph{{A Proof that the Free Energy of a Spin System is Extensive}}, J. Math. Phys. \textbf{5}, 1215 (1964)

\bibitem{Gorkov}
L.P. Gor'kov, \emph{{Microscopic derivation of the Ginzburg-Landau equations in
  the theory of superconductivity}}, Zh. Eksp. Teor. Fiz. \textbf{36} (1959),
  1918--1923, \emph{English translation} Soviet Phys.\ JETP \textbf{9},
  1364-1367 (1959).

\bibitem{HHSS08}
C.~Hainzl, E.~Hamza, R.~Seiringer, and J.P.~Solovej,
  \emph{{The BCS Functional for General Pair Interactions}}, Comm. math. phys. \textbf{281} (2008), no.~2, 349--367 .

\bibitem{HS13}
C.~Hainzl, B.~Schlein, 
\emph{{Dynamics of Bose-Einstein condensates of fermion pairs in the low density limit of BCS theory.}} 
J. Funct. Anal. \textbf{265} (2013), no. 3, 399-–423. 

\bibitem{HS12}
C.~Hainzl, R.~Seiringer, 
\emph{{Low density limit of BCS theory and Bose-Einstein condensation of fermion pairs.}}
Lett. Math. Phys. \textbf{100} (2012), no. 2, 119-–138

\bibitem{HSreview}
C.~Hainzl, R.~Seiringer, 
\emph{The Bardeen–Cooper–Schrieffer functional of superconductivity and its mathematical properties}, J. Math. Phys. 57, 021101 (2016)

\bibitem{HLP}
G.H.~Hardy, J.E.~Littlewood, and G.~Polya, 
\emph{Inequalities},
2nd ed. C.U.P. (1952)

\bibitem{Lipschitz}
S.~Hofmann, M.~Mitrea, and M.~Taylor, \emph{Geometric and transformational properties of Lipschitz domains, Semmes-Kenig-Toro domains, and other classes of finite perimeter domains}, J. Geom. Anal. 17 (2007), no. 4, 593–647

\bibitem{Leggett79}
A.J. Leggett, \emph{Diatomic molecules and cooper pairs}, Modern trends in the
  theory of condensed matter, J. Phys. (1980).

\bibitem{LiebSimonTF}
E.H.~Lieb and B.~Simon, 
\emph{{The Thomas-Fermi theory of atoms, molecules and solids}} Advances in Math. 23 (1977), no. 1, 22–116.
  

\bibitem{Marcusetal}
M.~Marcus, V.~Mizel, and Y.~Pinchover, \emph{{On the best constant for Hardy's inequality in $R^n$}} Trans. Amer. Math. Soc. \textbf{350} (1998), no. 8, 3237–3255

\bibitem{MS}
T.~Matskewich, and P.E.~Sobolevskii, 
\emph{The best possible constant in generalized Hardy's inequality for convex domain in $\R^n$}, Nonlinear Anal. 28 (1997), no. 9, 1601–1610.

\bibitem{Necas}
J.~Necas, \emph{Sur une m\'ethode pour r\'esoudre les \'equations aux d\'eriv\'ees partielles du type elliptique, voisine de la variationelle}, Ann. Scuola Norm. Sup. Pisa (3) 16 (1962), 305–-326

\bibitem{NS85}
P. Nozières, S. Schmitt-Rink, \emph{{Derivation of the Gross-Pitaevskii Equation for Condensed Bosons from the Bogoliubov–de Gennes Equations for Superfluid Fermions}}, Journal of Low Temperature Physics, \textbf{59} (3), 195--211 (1985)

\bibitem{PS03}
P. Pieri and G. C. Strinati, \emph{{Derivation of the Gross-Pitaevskii Equation for Condensed Bosons from the Bogoliubov–de Gennes Equations for Superfluid Fermions}}
Phys. Rev. Lett. 91, 030401 (2003)


\bibitem{SRE93}
C. A. R. Sá de Melo, M. Randeria, and J. R. Engelbrecht, \emph{{Crossover from BCS to Bose superconductivity: Transition temperature and time-dependent Ginzburg-Landau theory}}, Phys. Rev. Lett. \textbf{71}, 3202 (1993).

\bibitem{Simon}
B.~Simon, \emph{Schr\"odinger semigroups}, 
Bull. Amer. Math. Soc. (N.S.) \textbf{7} (1982), no. 3, 447–526. 

\bibitem{SteinSIDPF}
E.M.~Stein,
\emph{Singular integrals and differentiability properties of functions},  
Princeton Mathematical Series, No. 30 Princeton University Press, Princeton, N.J. (1970)


\end{thebibliography}

\end{document}